\documentclass[twocolumn]{aastex62}

\graphicspath{{./}}

\accepted{April 13, 2019}
\shorttitle{The GISMO 2mm survey in COSMOS}
\shortauthors{Magnelli et al.}
\begin{document}

\title{The IRAM/GISMO two-millimeter survey in the COSMOS field\footnote{Based on observations with the IRAM 30m-telescope.}}

\correspondingauthor{Benjamin Magnelli}
\email{magnelli@astro.uni-bonn.de}

\author[0000-0002-6777-6490]{B. Magnelli}
\affil{Argelander-Institut f\"ur Astronomie, Universit\"at Bonn, Auf dem H\"ugel 71, Bonn, D-53121, Germany}

\author[0000-0002-8414-9579]{A. Karim}
\affiliation{Argelander-Institut f\"ur Astronomie, Universit\"at Bonn, Auf dem H\"ugel 71, Bonn, D-53121, Germany}

\author[0000-0002-8437-0433]{J. Staguhn}
\affiliation{The Henry A. Rowland Department of Physics and Astronomy, Johns Hopkins University, 3400 North Charles Street, Baltimore, MD 21218, USA}
\affiliation{Observational Cosmology Lab, Code 665, NASA Goddard Space Flight Center, Greenbelt, MD 20771, USA}

\author[0000-0001-8991-9088]{A. Kov\'acs}
\affiliation{Smithsonian Astrophysical Observatory Submillimeter Array (SMA), MS-78, 60 Garden St, Cambridge, MA 02138, USA}

\author[0000-0002- 2640-5917]{E.F. Jim\'enez-Andrade}
\affiliation{Argelander-Institut f\"ur Astronomie, Universit\"at Bonn, Auf dem H\"ugel 71, Bonn, D-53121, Germany}

\author[0000-0002-0930-6466]{C.M. Casey}
\affiliation{Department of Astronomy, The University of Texas at Austin, 2515 Speedway Blvd Stop C1400, Austin, TX 78712, USA}

\author[0000-0002-7051-1100]{J.A. Zavala}
\affiliation{Department of Astronomy, The University of Texas at Austin, 2515 Speedway Blvd Stop C1400, Austin, TX 78712, USA}

\author[0000-0002-3933-7677]{E. Schinnerer}
\affiliation{Max Planck Institut f\"ur Astronomie, K\"onigstuhl 17, D-69117, Heidelberg, Germany}

\author[0000-0003-1033-9684]{M. Sargent}
\affiliation{Astronomy Centre, Department of Physics and Astronomy, University of Sussex, Brighton BN1 9QH, UK}

\author[0000-0002-6290-3198]{M. Aravena}
\affiliation{N\'ucleo de Astronom\'ia, Facultad de Ingenier\'ia, Universidad Diego Portales, Av. Ej\'ercito 441, Santiago, Chile}

\author[0000-0002-1707-1775]{F. Bertoldi}
\affiliation{Argelander-Institut f\"ur Astronomie, Universit\"at Bonn, Auf dem H\"ugel 71, Bonn, D-53121, Germany}

\author{P.L. Capak}
\affiliation{Infrared Processing and Analysis Center, California Institute of Technology, Pasadena, CA 91125, USA}

\author[0000-0001- 9585-1462]{D.A. Riechers}
\affiliation{Cornell University, Space Sciences Building, Ithaca, NY 14853, USA}
\affiliation{Max Planck Institut f\"ur Astronomie, K\"onigstuhl 17, D-69117, Heidelberg, Germany}

\author{D.J. Benford}
\affiliation{Astrophysics Division, NASA Headquarters, 300 E St. SW, Washington, DC 20546, USA}

\begin{abstract}
We present deep continuum observations at a wavelength of 2\,mm centered on the COSMOS field using the Goddard IRAM Superconducting Millimeter Observer (GISMO) at the IRAM 30\,m-telescope. 
These data constitute the widest deep 2\,mm survey to-date, reaching a uniform $\sigma$$\,\sim\,$$0.23$\,mJy\,beam$^{-1}$ sensitivity over $\sim\,$250\,arcmin$^2$ at $\sim\,$24\arcsec\ resolution.
We detect four sources at high significance (S/N$\,\geq\,$4.4) with a expected number of false detection of 0.09 sources, and five sources at 4.4$\,>\,$S/N$\,\geq\,$3.7 with a expected number of false detection of 1.65 sources.
Combined with deep GISMO observations in GOODS-N, we constrain the 2\,mm number counts over one decade in flux density.
These measurements agree with most galaxy evolution models tested here, except those with large population of dusty star-forming galaxies at $z>7$.
Five GISMO sources have counterparts in (sub-)millimeter catalogs available in COSMOS.
Their redshifts suggest that all but one lie above $z$$\,\sim\,$$3$.
These four high-redshift ($z$$\,>\,$$3$) galaxies have $\tilde{z}$$\,=\,$3.9, SFRs $\sim\,$400 -- 1200\,M$_{\odot}\,$yr$^{-1}$ and $M_{\rm dust}$$\,\sim10^{9.5}\,$M$_{\odot}$.
They provide a relatively complete selection ($\sim$$\,66\%$) of the most luminous ($L_{\rm IR}$$>\,$10$^{12.6}\,$L$_{\odot}$) and highest redshift ($z>3$) galaxies detected within our survey area by AzTEC at 1.1\,mm.
We thus conclude that 2\,mm surveys favor the selection of massive, vigorously star-forming, high-redshift galaxies.
This is corroborated by GISMO-C4, a source with a low false detection probability ($\sim\,$6.2\%), for which the absence of a (sub-)millimeter counterpart supports a high redshift origin ($z\gtrsim3$).
\end{abstract}


\section{Introduction}
\label{sec:intro}
One of the most pressing questions in extragalactic astronomy concerns the production of dust in the very early universe.
In particular, the amount of dust that could have reasonably been formed in primordial galaxies within the first few hundred million years after the Big Bang is still heavily debated theoretically and also observationally \citep[e.g.][]{michalowski_2015,mancini_2015,wang_2017}. 
Indeed, dust formation in asymptotic giant branch (AGB) star atmospheres is expected to take too long ($\gtrsim$\,500\,Myr) to explain the large dust reservoir observed in some $z$$\,\gtrsim\,$$5$ galaxies; while a significant fraction of the dust formed on short timescales ($<$\,500\,Myr) by supernovae (SNe) is expected to be destroyed by the associated shocks.
Unfortunately, to-date only a few objects could be studied thoroughly in their cold dust properties for a meaningful estimate of their individual dust masses, including bolometrically bright quasar host galaxies, $\gamma$-ray burst host galaxies, or gravitationally lensed galaxies at redshifts $z$$\,\gtrsim\,$$5$ \citep[e.g.][]{michalowski_2010,riechers_2013,hjorth_2013,michalowski_2015,strandet_2017,zafar_2018,marrone_2018,decarli_2018,venemans_2018}.
Such observational constraints do not, however, qualify for robust extrapolation to the total dust mass in the early universe, as these galaxies are not necessarily representative of the broader population. 

Despite some limitations, work over the past two decades has led to the discovery of considerable populations of dust-rich star-forming galaxies (DSFGs) at high redshifts (though largely found below $z$$\,=\,$$4$), through deep field survey campaigns at (sub-)millimeter (hereafter (sub)mm) wavelengths \citep*[e.g., see reviews of][]{blain_2002,casey_2014}. 
Surveys of such DSFGs benefit from a strongly negative $K$-correction at (sub)mm wavelengths, which renders galaxies of equal total infrared luminosity and cold dust temperature equally likely to be detected in a flux-limited survey, no matter if they reside at $z\sim1$ or $z\sim10$. 
This effect is due to the steep slope of the Rayleigh-Jeans continuum emitted by interstellar dust grains heated by the radiation field of young stars in star-forming galaxies.  
Unfortunately, distinguishing in (sub)mm surveys the dustiest galaxies at the highest redshifts ($z>4$) from lower-redshift galaxies is exceedingly challenging due to the difficulty in making spectroscopic identifications at those epochs; or is observationnally expensive if one used multiple (sub)mm color criteria such as the \textit{Herschel}-Red sources \citep[e.g.,][]{cox_2011,riechers_2013,ivison_2016}.
In addition, current (sub)mm surveys being mostly performed at $850$$\,\mu$m$\,<\,$$\lambda_{\rm obs}<1.3\,$mm, at the highest redshifts ($z>4$) they do not probe sufficiently long rest-frame wavelengths to yield accurate dust mass estimates \citep[i.e., $\lambda_{\rm rest}\gtrsim250\,\mu$m;][]{scoville_2016}.

In principle, the 2\,mm window provides an ideal setup for deep field surveys aiming at statistical probes of dust-rich objects at these early epochs.
It probes sufficiently long rest-frame wavelengths for accurate dust mass estimates, while ``filtering out'' lower redshift dusty sources at $z\sim2$, which ``contaminate'' the 850\,$\mu$m and 1.3\,mm surveys \citep{casey_2018,casey_2018b}.
Unfortunately, there have not been to-date the observational datasets to test this latter hypothesis.
Indeed, until recently the 2\,mm window has mainly been explored by large-scale surveys, such as those performed by the South Pole Telescope \citep[SPT;][]{vieira_2010}, with detection limits of few milli-Janskys and thus restricted to the study of gravitationally-lensed galaxies.
     
Here we present a pioneering survey in this largely unexplored 2\,mm window using the Goddard IRAM Superconducting Millimeter Observer (GISMO) as guest instrument at the IRAM 30m-telescope \citep[][hereafter S14]{staguhn_2014}.
In order to complement the deep pencil-beam observations with GISMO towards the northern Great Observatories Origins Survey (GOODS-N; described in S14) field, we targeted a $\sim\,$$0.1\,$deg$^2$ area towards the central portion of the equatorial 2\,deg$^2$ COSMOS field \citep{scoville_2007}.
Our GISMO 2\,mm survey covers a part of the COSMOS field with the deepest multi-wavelength coverage, including Hubble/WFC3 data from the CANDELS survey \citep{koekemoer_2011}.
It overlaps with all (sub)mm surveys undertaken in COSMOS, including deep AzTEC/JCMT \citep{scott_2008} and AzTEC/ASTE \citep{aretxaga_2011} 1.1\,mm surveys and {\sc Scuba}-2/JCMT 450 and 850\,$\mu$m surveys \citep{casey_2013,geach_2017,wang_2017dr}.
In addition, sub-millimeter galaxies (SMGs) selected from these surveys benefited from numerous interferometric follow-ups \citep[SMA/CARMA/PdBI/ALMA;][]{younger_2007,younger_2009,aravena_2010,smolcic_2012,brisbin_2017}, offering precise locations and redshift estimates for a large fraction of them \citep{brisbin_2017,miettinen_2017}, including some distant ($z$$\,>\,$4) starbursts, a sub-class of SMGs identified only in recent years \citep{smolcic_2012,riechers_2014,smolcic_2015,miettinen_2015,brisbin_2017}.
This wealth of deep pan-chromatic ancillary data available within our GISMO survey is vital to test the hypothesis that the 2\,mm window unveils the dustiest and most distant starbursts of the Universe.

The focus of this paper is two-fold.
While we will analyze the individual detections with respect to their redshift and multi-wavelength properties, we will thoroughly discuss statistical implications arising from a novel constraint of the bright-end of the 2\,mm source counts.
The latter have been previously predicted by several authors using different semi-empirical techniques \citep[][]{bethermin_2017,casey_2018,zavala_2018} and are supposed to be a useful tool to distinguish between different model universes -- from the very dust-rich to the very dust-poor \citep{casey_2018}.
Providing an observational constraint on the 2\,mm source counts is thus paramount to elaborate on the best future strategies to statistically explore the interstellar medium at the earliest cosmic epochs.
Throughout this paper we assume a \textit{Planck} cosmology, adopting $H_{0}=67.8$~(km/s)/Mpc, $\Omega_M=0.308$ and $\Omega_{\Lambda}=0.692$ \citep{planck_cosmo_2016}. 
A \citet{chabrier_2003} initial mass function (IMF) is used for all stellar mass and SFR measurements in this article.

\section{The COSMOS-GISMO 2\,mm survey}
\subsection{Observations}
\label{subsec:obs}
The GISMO observations of the COSMOS field were obtained in pooled campaigns over the course of three years (April 2012, April 2013, April 2014, October 2014 and February 2015)\footnote{The corresponding IRAM project IDs are 247-11, 227-12, 242-13, 117-14 and 232-15.} at the IRAM 30-m telescope as a combined open/guaranteed time program for a total of 113.6\,hrs (incl. $\sim\,$37\%\ calibration/instrumental overheads, with a total of $\sim\,$71\,hrs on target).

GISMO consists of $8 \times16$ pixels \citep{staguhn_2008} with super conducting transition edge sensors (TES). 
The TES are read out by time domain SQUID multiplexers built at the National Institute for Standards (NIST), in Boulder, Colorado \citep{irwin_2002}. 
The GISMO bandpass has a full width at half maximum (FWHM) of $\sim\,$25\,GHz around its peak at 150\,GHz (i.e., 2\,mm). 
Pixels are spaced by $13\farcs75$ and the instantaneous field-of-view is $1\farcm8$$\,\times\,$$3\farcm7$.
More details about the instrument can be found in S14.

The beamsize of GISMO observations at the 30-m telescope is $16\farcs6$ FWHM.
The observations were generally carried out under stable atmospheric conditions and $<\,$4\,mm perceptible water vapor, in other words, with a zenith opacity of $\tau_{\rm{2mm}}$$\,<\,$$0.11$.
On average we found 100 pixels to be working during our observations. 
Focus in $z$-direction was regularly monitored (four times a day) and pointing was frequently checked (once per observing hour) using the nearby bright quasars J1055+018, J0851+202 and J0823+033. 
Fluxes were calibrated to $<\,$10\% accuracy by monitoring Mars, Uranus, and Neptune and employing the atmospheric transmission model of the Caltech Submillimeter Observatory\footnote{\url{http://www.submm.caltech.edu/cso/weather/atplot.shtml}} and the 30-m telescope 225\,GHz radiometer readings. 
We employed an on-the-fly raster scan pattern comprised of 41 $\times$ 10--11s (depending on target elevation) subscans and additional 3s turnaround overhead to cover a total square area of $20\arcmin$$\,\times\,$$20\arcmin$ per full scan. 
This observing mode works in total power, hence the signal is not modulated by switching the secondary mirror. 
Per subscan, additional signal was obtained over an area corresponding to the array extent. 
The short subscan duration resulted in fast scanning speeds between 110--120\arcsec\ per second and was chosen to minimize the impact of $1/f$ noise on our large map. 
We changed the scanning direction by 45\,deg about the current azimuth between every two scans in order to assure a homogeneous coverage of the target area and reduce systematic effects in the data. 
We typically observed for contiguous 7--8\,hr blocks and hence profited from earth rotation to further enhance a homogeneous target coverage. 
Consequently, the resulting multi-season map is circular and reaches a uniform $\sigma$$\,\sim\,$$0.23$ ($0.3$) mJy/beam sensitivity over an area with $\sim\,$17\farcm8 (21\farcm4) diameter, i.e. $\sim\,$0.07 (0.1) deg$^2$.
At the radius of the full map (i.e., $12\arcmin$) around the central coordinates ($\alpha=10^{\rm h}\,00^{\rm m}\,19\fs75,\ \delta=+02\arcdeg\,32\arcmin\,04\farcs40$; J2000) a $\sigma$$\,\sim\,$$0.4\,$mJy/beam sensitivity is reached. 

\subsection{Data Reduction}
\label{subsec:reduction}
The data have been reduced using version 2.30-4 of the \texttt{CRUSH} software \citep{kovacs_2008} in deep mode used for the accurate recovery of point sources (for details see S14). 
In this process the map is spatially filtered above 60\arcsec\ FWHM to remove spatially variant atmospheric residuals and smoothed with a Gaussian FWHM of 17\farcs5 (i.e., $\sim$matched-filtered) to optimize the detection of point sources.
The resulting effective image resolution is 24\arcsec\ FWHM.
\begin{figure}
	\begin{center}
	\includegraphics[width=\linewidth]{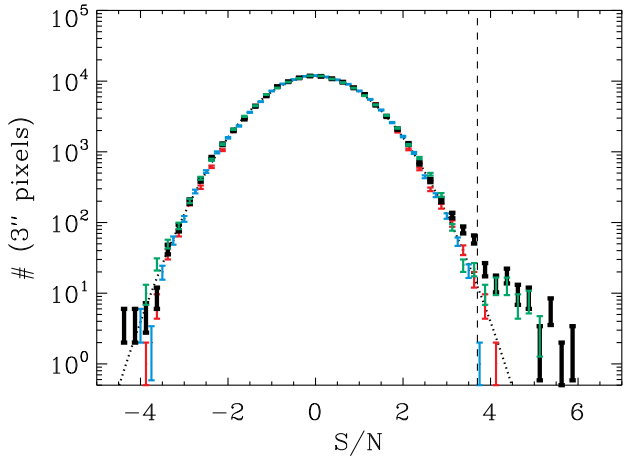} 
	\caption{\label{fig:pixel_dist}
    Signal-to-noise histogram for the smoothed and filtered map (black), for a jackknifed realization (red) and expectation for a Gaussian noise distribution with $\sigma=1$ (dotted line). 
    As demonstrated by the good agreement between these three histograms, noise in our map is Gaussian in nature and does not include a significant contribution from unresolved sources (i.e., confusion noise).
    The asymmetric excess on the positive half of the distribution for the smoothed and filtered map comes from resolved sources.
    This asymmetric excess is, as expected, absent in the histogram for the residual map (blue), i.e., the smoothed and filtered map after sources detected with $\rm S/N\geq3.7$ (vertical dashed line; see Sect. \ref{subsec:extraction}) are removed.}
    \end{center}
\end{figure}
\begin{figure*}
	\begin{center}
	\includegraphics[width=0.73\linewidth]{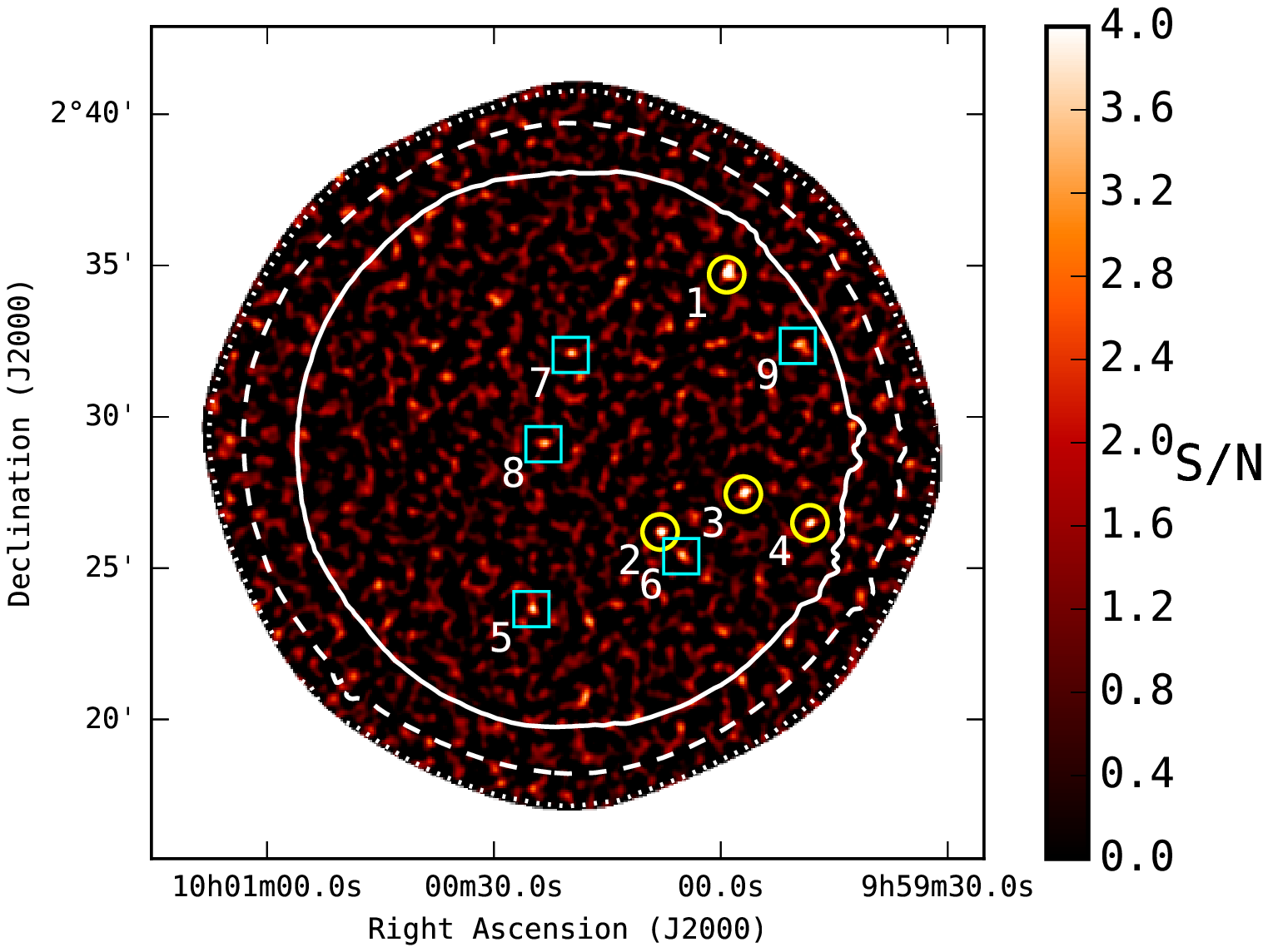} 
	\caption{\label{fig:GISMOmap}
    The signal-to-noise ratio map for the smoothed and filtered GISMO COSMOS field. 
    The solid, dashed and dotted contours encompass areas of uniform 0.23, 0.30 and 0.4\, $\rm mJy \,beam^{-1}$ rms, respectively.
    Circles show sources detected at S/N$\,\geq\,$4.4 (Tab.~\ref{tab:sample}), where the expected number of false detection is $\sim$0.09 sources.
    Squares show sources detected at $\rm 4.4>S/N\geq3.7$ (Tab.~\ref{tab:sample}), where the expected number of false detection is $\sim$1.65 sources.}
    \end{center}
\end{figure*}

To estimate the noise, we produced ``jackknifed'' maps with \texttt{CRUSH} by randomly multiplying each scan by $+1$ or $-1$, eliminating any stationary noise (including sources and foregrounds) but retaining random noise (including that from the atmosphere).
From these jackknifed realizations, we generated the noise map associated with the regular smoothed and filtered map.
Fig.~\ref{fig:pixel_dist} shows the histogram of the signal-to-noise ratio (S/N) of one of these jackknifed, filtered and smoothed realization compared to the S/N histogram of the regular filtered and smoothed map.
The jackknifed distribution is well fit by a Gaussian with $\sigma=1$, demonstrating that noise in the GISMO map is indeed Gaussian.
The histogram for the regular map is also well described by this Gaussian, except at high S/N where resolved sources create an asymmetric excess.
This agreement demonstrates that the regular map does not contain a significant stationary noise contribution from unresolved sources (i.e., confusion noise) that would appear as a symmetric widening of the distribution with respect to the jackknifed expectation.
This is in clear contrast to the GISMO Deep Field (GDF; S14) in which a significant fraction of the total noise comes from confusion.
Indeed, in the GDF, the integration time per beam is long enough for the random noise ($\sigma_{\rm rand.}\propto t^{-1/2}$) and the confusion noise ($\sigma_{\rm confusion}$) to be comparable and thus for a significant fraction of the total noise ($\sigma_{\rm tot}$) to come from confusion ($\sigma_{\rm tot}^2=\sigma_{\rm rand.}^2+\sigma_{\rm confusion}^2$).
In our COSMOS survey, $\sigma_{\rm rand.}$ dominates and $\sigma_{\rm confusion}$  can be neglected.
Note that the subtraction of sources truncates the S/N histogram for the residual map at our $3.7$$\,\sigma$ detection threshold (see Sect.~\ref{subsec:extraction}).

\subsection{Source Extraction}
\label{subsec:extraction}
Source extraction was performed on the smoothed and filtered signal and noise maps produced by \texttt{CRUSH}. 
The value of these beam-smoothed maps provides the amplitude and uncertainty of a fitted point spread function (PSF) at each position, which in case of point sources corresponds directly to their fluxes and associated uncertainties (see S14).
To extract these fluxes, we employed a standard top-down peak-finding approach:
(\textit{i}) we search for the highest peak in the signal-to-noise ratio map; (\textit{ii}) the flux, noise and position of this source is cataloged; (\textit{iii}) it is removed from the smoothed and filtered map using a scaled version of our effective PSF (see below) and a new signal-to-noise ratio map is generated; (\textit{iv}) we repeat this procedure until the highest peak in the signal-to-noise ratio map is below 3$\sigma$, i.e., a level under which sources are deemed to be unreliable. 
The effective PSF used in this process was generated by convolving the instrumental 16\farcs6 FWHM Gaussian beam with a 17\farcs5 FWHM Gaussian ($\sim$matched-filtering) and a negative 60\arcsec\ FWHM Gaussian bowl (resulting from our spatial-filtering above 60\arcsec\ FWHM; Sect.~\ref{subsec:reduction}).
The resulting PSF has a peak of one, zero integral (no DC sensitivity due to filtering; S14) and an 24\arcsec\ FWHM.

In Sect.~\ref{subsubsec:contanimation}, we further refined the S/N threshold above which sources are considered as reliable.
We cut our final catalog at 3.7$\sigma$, where the \textit{overall} false-detection rate is 20\%, but particularly highlight sources above 4.4$\sigma$ where the \textit{overall} false-detection rate is only 2.5\%.
A total of nine sources were found with S/N$\geq3.7$, among which four have S/N$\geq4.4$, within the uniform $\sigma$$\,\sim\,$$0.23$ mJy/beam sensitivity area of our map (Fig.~\ref{fig:GISMOmap}; Tab.~\ref{tab:sample}).
In the rest of the paper, we strictly limit our analysis to this homogeneously deep part of the map, which covers $\sim\,$250\,arcmin$^2$.

\begin{table*}
\begin{center}
\caption{\label{tab:sample} GISMO-Detected Source List}
\begin{tabular}{ c c c c c c c c c } 
\hline \hline
\rule{0pt}{3ex}ID & R.A. & Decl. & S/N & $S_{\rm 2mm}$ & \multicolumn{3}{c}{$S_{\rm 2mm}$} & $P_{\rm f}$$^{\rm b}$ \\
   &      &       &     & raw & \multicolumn{3}{c}{deboosted$^{\rm a}$}  \\ 
\cline{6-8}
   &      &       &     &     &  \rule{0pt}{3ex}Zavala+19 & B\'ethermin+17 & Staguhn+14 \\
  & (J2000)& (J2000) & & (mJy) & (mJy) & (mJy) & (mJy) & (\%) \\
\hline
\rule{0pt}{3ex}GISMO-C1 & 09:59:59.2   &    +2:34:41.89   &    5.8  &     1.29 $\pm$ 0.22   &     $1.11^{+0.20}_{-0.31}$ & $1.07^{+0.20}_{-0.31}$ & $1.15^{+0.25}_{-0.25}$ & 0.0 \\
\rule{0pt}{3ex}GISMO-C2 & 10:00:08.0   &    +2:26:11.90   &    5.0   &     1.09 $\pm$ 0.22   &     $0.90^{+0.18}_{-0.35}$ & $0.80^{+0.19}_{-0.33}$ & $0.90^{+0.27}_{-0.27}$ & 1.4 \\
\rule{0pt}{3ex}GISMO-C3 & 09:59:57.0   &    +2:27:26.89   &    4.6  &     1.02 $\pm$ 0.22   &     $0.81^{+0.18}_{-0.35}$ & $0.69^{+0.20}_{-0.32}$ & $0.80^{+0.28}_{-0.28}$ & 4.0  \\
\rule{0pt}{3ex}\rule{0pt}{-3ex}GISMO-C4 & 9:59:48.2   &    +2:26:29.88   &    4.4  &     1.01 $\pm$ 0.23   &     $0.76^{+0.21}_{-0.35}$ & $0.65^{+0.20}_{-0.34}$ & $0.75^{+0.30}_{-0.30}$ & 6.2 \\
\hline
\rule{0pt}{3ex}GISMO-C5 & 10:00:25.0   &    +2:23:38.90   &    3.9  &     0.86 $\pm$ 0.22   &     $0.55^{+0.22}_{-0.35}$ & $0.48^{+0.13}_{-0.33}$ & $0.55^{+0.30}_{-0.30}$ & 23.3 \\
\rule{0pt}{3ex}GISMO-C6 & 10:00:05.2   &    +2:25:23.90   &    3.9  &     0.84 $\pm$ 0.22   &     $0.53^{+0.21}_{-0.35}$ & $0.47^{+0.13}_{-0.32}$ & $0.52^{+0.30}_{-0.30}$ & 24.1 \\
\rule{0pt}{3ex}GISMO-C7 & 10:00:19.8   &    +2:32:02.90   &    3.8  &     0.84 $\pm$ 0.22   &     $0.53^{+0.21}_{-0.35}$ & $0.47^{+0.13}_{-0.32}$ & $0.52^{+0.30}_{-0.30}$ & 24.1 \\
\rule{0pt}{3ex}GISMO-C8 & 10:00:23.4   &    +2:29:05.90   &    3.7  &     0.82 $\pm$ 0.22   &     $0.50^{+0.19}_{-0.35}$ & $0.43^{+0.13}_{-0.31}$ & $0.49^{+0.29}_{-0.29}$ & 31.7 \\
\rule{0pt}{3ex}GISMO-C9 & 09:59:49.8   &    +2:32:20.88   &    3.7  &     0.82 $\pm$ 0.22   &     $0.50^{+0.19}_{-0.35}$ & $0.43^{+0.13}_{-0.31}$ & $0.49^{+0.29}_{-0.29}$ & 31.7 \\
\hline
\end{tabular}
\end{center}
\textbf{Notes.}
$^{\rm{(a)}}$ Deboosted fluxes estimated using the Monte-Carlo approach described in Sect.~\ref{subsubsec:flux boosting} based on the number counts of \citet{zavala_2018} and \citet{bethermin_2017}, while in the rightmost column the deboosted fluxes were measured using the analytical approach described in \citet{staguhn_2014} based on the number counts of \citet{bethermin_2011}.
$^{\rm{(b)}}$ False-detection rates at S/N$\,\pm\,\Delta$S/N, defined as the mean false-detection rates inferred from the models of \citet{zavala_2018} and \citet{bethermin_2017}.
The separation at S/N$\,\geq\,$4.4 corresponds to the detection significance above which the \textit{overall} false-detection rate is only 2.5\%.
We cut our final catalog at 3.7$\sigma$, above which the \textit{overall} false-detection rate is 20\%.
\end{table*}

\subsubsection{Flux boosting}
\label{subsubsec:flux boosting}
Low angular resolution maps in the millimeter have the potential to suffer from a set of boosting effects which result in measured fluxes for sources higher than their intrinsic values. Flux boosting can be caused by Eddington bias (in the event that the intrinsic number counts at the observed wavelength are steep) and confusion noise (the high density of sources per beam boosts the measured flux density of any single measured source).
To evaluate this flux boosting as a function of the \textit{observed} signal-to-noise ratio (where the signal is the measured ``raw'' flux density), we performed extensive Monte-Carlo simulations.
We inserted fake sources following a realistic number count distribution into our jackknifed maps and recovered their positions, fluxes and flux uncertainties applying the same source extraction method as that used for our real map.
\begin{figure}
\includegraphics[width=\linewidth]{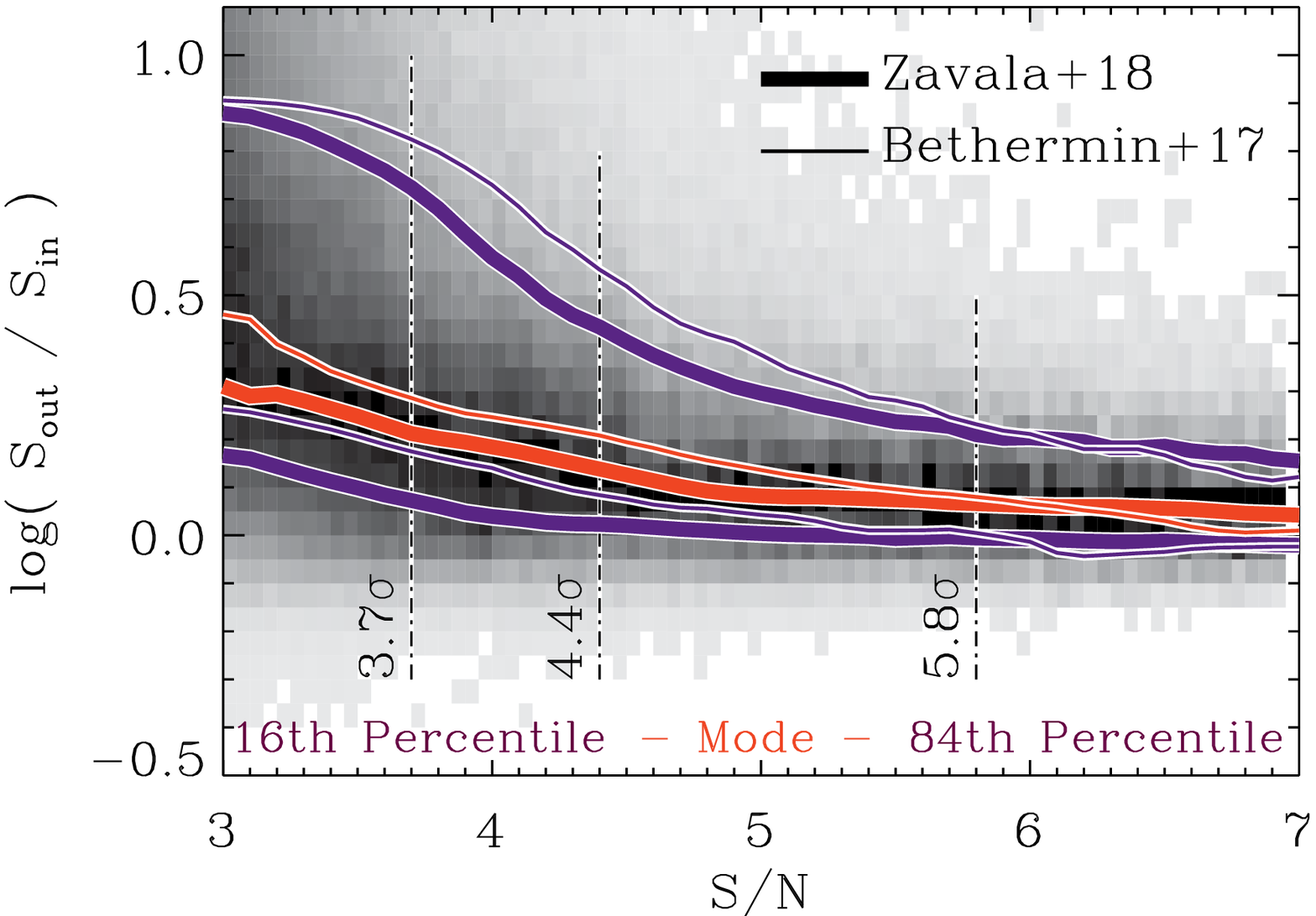}		
\caption{\label{fig:sout_sin}
		Flux boosting as a function of \textit{observed} signal-to-noise ratio estimated from simulations. 
        The density distribution of data points from simulations following the model of \citet{zavala_2018} is shown by the shaded region.
        For clarity, shadings are independent for each signal-to-noise ratio bin, i.e., the darkest color indicates the highest number density of data points in this signal-to-noise ratio bin. 
        The 16th percentile (blue), mode (red) and 84th percentile (blue) of the distribution as a function of S/N are shown by the thick lines.
        For comparison, the 16th percentile, mode and 84th percentile of the distribution inferred from the \citet{bethermin_2017} simulations are shown by the thin lines.
        In the range probed by real detections (i.e., 5.8$\,\geq\,$S/N$\,\geq\,$3.7; vertical dot-dashed line), these models yield slightly different flux boosting statistic, yet consistent within their uncertainties.
        }
\end{figure}

Here, we adopted two different number count distribution models, that of \citet{zavala_2018} and \citet{bethermin_2017}.
The \citet{bethermin_2017} SIDES model uses an halo-occupation distribution modeling and infrared spectral energy distributions (SEDs) drawn from a two star-formation modes galaxy evolution model \citep{bethermin_2012b}; while the \citet{zavala_2018} model infers the number counts at 2\,mm using the first 3\,mm number count measurements combined with the infrared luminosity function models discussed in \citet{casey_2018} and \citet{casey_2018b}.
These models have slightly different number count distributions as a result of  different luminosity function assumptions at $z>4$.
This give us the ability to explore how those assumptions affect the flux boosting statistic.
It is worth noting, however, that {\it both} models agree within the uncertainties with the number counts  from our measurements (see Sect.~\ref{subsec:counts}).
Injected sources of both models have fluxes as low as $0.05\,$mJy (i.e., well below our detection threshold) and are randomly positioned into our jackknifed maps.
We verified that the signal-to-noise histogram of these simulated maps peaks at zero and has a dispersion of one, similarly as our real map (see Fig.~\ref{fig:pixel_dist}).
Sources recovered in these simulated maps above 3$\sigma$ are associated with the brightest input source within a radius of 18\arcsec.
This radius corresponds to the GISMO astrometric accuracy as inferred from Eq.~\ref{eq:position} using S/N$\,=\,$3 (Sect.~\ref{subsubsec:position}; see also Eq.~9 of S14).
For each number count model, we generated 10,000 mock maps, recovering about half a million fake sources above 3$\sigma$.

The output-to-input flux ratio (i.e., $S_{\rm out}/S_{\rm in}$) as a function of the \textit{observed} signal-to-noise ratio (i.e., $S_{\rm out}$/N) inferred from our simulations is shown in Fig.~\ref{fig:sout_sin}.
While at high S/N flux boosting converges towards zero and thereby becomes insignificant, it steadily increases at lower S/N, reaching up to $\sim$0.3\,dex at 3.7$\,\sigma$.
In the S/N range probed by real detections (i.e., 5.8$\,\geq\,$S/N$\,\geq\,$3.7), the model of \citet{zavala_2018} predicts lower flux boosting statistics ($\sim\,$0.05\,dex), as its number count distribution has a shallower slope in the corresponding flux density range than that of \citet{bethermin_2017} (see Sect.~\ref{subsec:counts}).
These differences are, however, well within the uncertainties of each model and their effects on the measured number counts are further discussed in Sect.~\ref{subsec:counts}.

In Tab.~\ref{tab:sample}, we tabulate the ``deboosted'' fluxes and uncertainties of our real detections, as inferred from both models. 
Deboosted fluxes correspond to the mode of the distributions at a given S/N, while the upper and lower uncertainties correspond to their 16th and 86th percentile, respectively.
We also tabulated in Tab.~\ref{tab:sample} deboosted fluxes and uncertainties obtained using the analytical approach described in S14 (see their Sect.~4.2) and assuming the number count distribution of \citet{bethermin_2011}.
These deboosted fluxes are fully consistent within their uncertainties with those inferred here. 

\subsubsection{Completeness}
\begin{figure}
	\begin{center}
	\includegraphics[width=\linewidth]{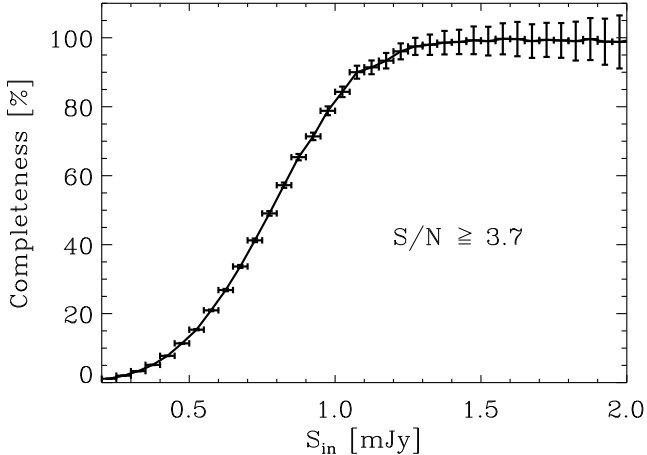}
	\caption{\label{fig:Completeness}
    		Completeness derived from the fraction of simulated sources injected into our jackknifed maps that are recovered by the source extraction at a significance $\geq\,$3.7$\,\sigma$ (irrespective of their flux accuracies) as a function of input flux.
            Errors bars correspond to the 1-$\sigma$ Poisson uncertainties measured using the number of sources available to assess the completeness within a given input flux bin.
    		}
    \end{center}
\end{figure}

We also used our simulations to evaluate the source detection completeness of our final catalog.
The completeness was defined as the ratio of recovered sources above a given S/N to the total number of input sources at a given flux density.
Completeness was evaluated in bins of \textit{input} flux density and for a detection significance $\geq\,$3.7$\,\sigma$, which was the final cut of our catalog described in the next subsection.
The completeness was evaluated for both simulations but only that inferred from the model of \citet{zavala_2018} is shown in Fig.~\ref{fig:Completeness}. 
Indeed, because the completeness does not depend on the input number count distribution but simply on the map noise properties, both models yield exactly the same source detection completeness function.

The completeness is close to zero at low input fluxes (i.e., $\sim\,$0.3\,mJy) but increases rapidly, reaching 40\% and 80\% at $\sim\,$0.7\,mJy and $\sim\,$1.0\,mJy, respectively.
The completeness shown in Figure~\ref{fig:Completeness} is used in Sect.~\ref{subsec:counts} to measure the number counts.

\subsubsection{False-detection rate}
\label{subsubsec:contanimation}
\begin{figure}
	\begin{center}
	\includegraphics[width=\linewidth]{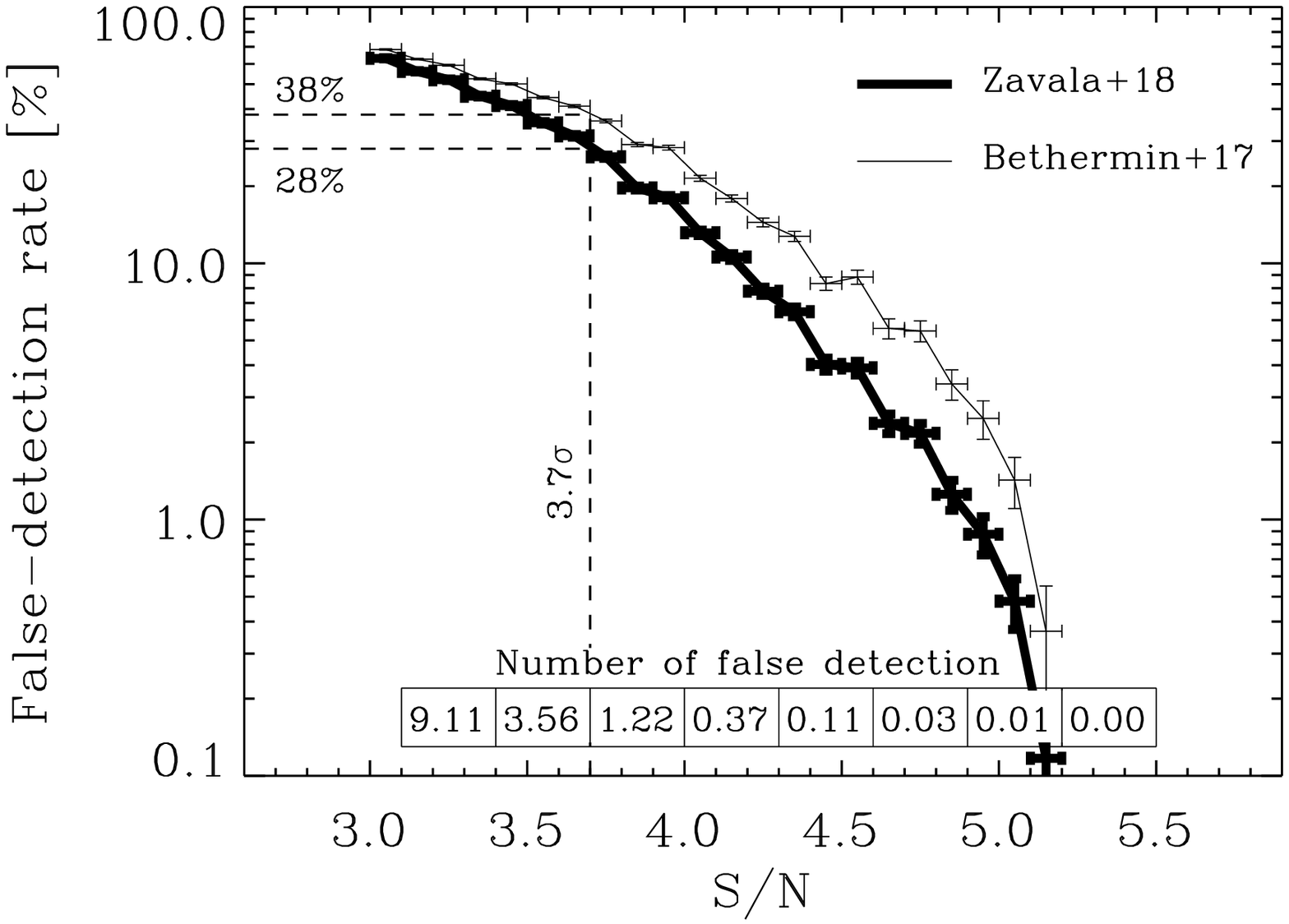}
	\caption{\label{fig:spurious}
    		False-detection rate derived from the ratio of sources recovered in the pure noise jackknifed maps to that recovered in our mock maps as a function of the \textit{observed} signal-to-noise ratio.
            Thick and thin lines are for sources recovered in mock maps following number count distributions as in \citet{zavala_2018} and \citet{bethermin_2017} models, respectively.
            Errors bars correspond to 1-$\sigma$ Poisson uncertainties measured using the number of sources available to assess the false-detection rate within a given S/N bin.
            Inset numbers correspond to the average number of false detection per map found in our pure noise jackknifed maps within a significance (i.e., S/N) range given by the lateral box outlines.
            While the numbers of false detections remain the same for both models, the number of sources recovered in their respective mock maps depends on the input number count distribution.
			Therefore, the models of \citet{zavala_2018} and \citet{bethermin_2017} yield slightly different false-detection rates.
			Among the five sources detected in the real map at 4.4$\,>\,$S/N$\,\geq\,$3.7, we expect 1.65 false detections.
            Among the four sources detected in the real map with S/N$\,\geq\,$4.4, we expect only 0.09 false detections.
    		}
    \end{center}
\end{figure}

The number of false detections in our map corresponds to the number of pure noise fluctuations reaching a given significance (i.e., S/N) and thus are mistakenly identified as real sources.
In the case of Gaussian noise fluctuations, this number depends only on the considered significance and number of independent Gaussian variables in the map.
However, because the number of independent Gaussian variables cannot be trivially calculated for our smoothed and filtered map, we evaluated the expected number of false detection using an empirical approach.
We run our source extraction algorithm on 10,000 pure jackknifed maps -- which by definition contain only noise -- and measured the average number of false detection recovered per map in bins of signal-to-noise ratio (see inset table of false sources identified per map in Fig.~\ref{fig:spurious}).
At S/N$\,\geq\,$4.4, where our real catalog contains four sources, the expected number of false detection is relatively low and equal to 0.09, corresponding to an \textit{overall} false-detection rate of $2.5$\%.
However, at lower significance, the number of false detection rises rapidly, and, for example, no less than 13.74 false detections are expected with S/N$\,\geq\,$3.1. 
In the real map, we recovered only 27 sources with S/N$\,\geq\,$3.1, leading to an unacceptably high \textit{overall} false-detection rate of about 51\%.
We thus decided to cut our final catalog at 3.7$\,\sigma$, where the \textit{overall} false-detection rate is at an acceptable value of 20\%, i.e., 1.74 false detections are expected with S/N$\,\geq\,$3.7 while we detected nine sources at this significance in the real map. 
Note that the number of false detections predicted by Eq.~2 of S14 is in very good agreement with our empirical estimates.

While measuring number counts, a false detection rate correction needs to be applied to each source according to its detection significance (see Sect.~\ref{subsec:counts}). 
Unfortunately, the low number of sources detected in the map does not allow us to infer a false-detection rate at any given significance (i.e., S/N$\,\pm\,\Delta$S/N) without being considerably affected by low number statistics.
Therefore, we evaluated these false-detection rates from our simulations, dividing the number of false detections recovered at a given significance (i.e., S/N$\,\pm\,\Delta$S/N) in the 10,000 pure jackknifed maps by the number of sources recovered at this significance in the 10,000 simulated maps of a given model (Fig.~\ref{fig:spurious}).
While the numbers of false detections remain the same for either models, the numbers of sources recovered in the simulated maps depend on the input number count distribution.
Therefore, the two models yield slightly different false-detection rates. 
At 3.7$\,\sigma$, the false-detection rate measured from the models of \citet{zavala_2018} and \citet{bethermin_2017} are 28\% and 38\%, respectively, in good agreement with that obtained by dividing the number of false detection with the number of sources recovered in the real map at 4.4$\,>\,$S/N$\,\geq\,$3.7, i.e., 1.65 false detections for five sources in the real map (i.e., 33\%).
Note that in Sect.~\ref{subsec:counts}, the false-detection rate of each model is used when appropriate, but in Tab.~\ref{tab:sample} we tabulated, for each source, the mean false-detection rate from these two models.

\subsubsection{Positional uncertainties}
\label{subsubsec:position}
\begin{figure}
	\begin{center}
	\includegraphics[width=\linewidth]{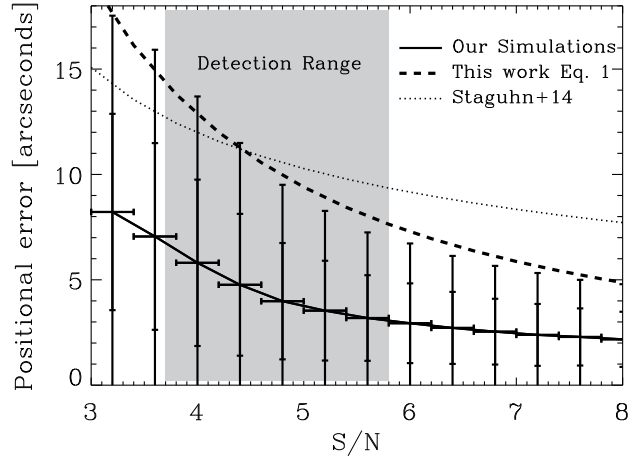}
	\caption{\label{fig:position}
    		Mean, 1$\,\sigma$ and 2$\,\sigma$ position uncertainties in bins of signal-to-noise ratio for source recovered in simulations following the model of \citet{zavala_2018}.
            Our 2$\,\sigma$ positional uncertainties are compared to predictions from Eq. 9 of S14 (dotted lines), assuming $\sigma_{\rm p}=0$ as there are no intrinsic position uncertainties in our input catalog.
            These 2$\,\sigma$ position uncertainties are fitted by a simple power law (dashed line; see Eq.~\ref{eq:position}).
            The shaded area shows the range of S/N probed by our real detections.
            The last bin is affected by our maximum matching radius of 18\arcsec, and thus likely underestimated.}
    \end{center}
\end{figure}

Finally, we evaluated the positional uncertainties of our catalog as the difference between the input and recovered positions of injected sources in our simulations.
Figure \ref{fig:position} shows the mean, 1$\,\sigma$  and 2$\,\sigma$ positional uncertainties of sources recovered from simulations following the model of \citet{zavala_2018} in bins of \textit{observed} signal-to-noise ratio. 
Note that the last signal-to-noise ratio bin is affected by our maximum matching radius of 18\arcsec\ (see Sect.~\ref{subsubsec:flux boosting}), and thus likely underestimated. 
As estimates from the model of \citet{bethermin_2017} are very similar, we did not show them in Fig.~\ref{fig:position}.

As noted in, e.g., \citet{ivison_2007}, S14 and \citet{geach_2017}, the positional uncertainties of point-like sources vary with their detection significance (i.e., S/N).
S14 defined in their Eq.~9 the maximum allowable separation between a GISMO source and that from other catalogs, taking into account the GISMO positional uncertainties and the 1$\,\sigma_{\rm p}$ catalog position errors.
In Fig.~\ref{fig:position}, we plotted their predictions, setting $\sigma_{\rm p}$$\,=\,$$0$ as there are naturally no intrinsic position errors in our simulated catalog.
These predictions should be compared to our 2$\,\sigma$ positional uncertainties, as they correspond to the radius where the counterpart must fall with $\sim$98\% confidence.
While in broad agreement, these predictions do not perfectly capture the trend with S/N observed in our simulations. 
We thus fitted these positional uncertainties with a simple power law, parametrized with the signal-to-noise ratio (SNR):
\begin{equation}
\label{eq:position}
\Delta \alpha = 9\farcs4\times \bigg( \frac{{\rm SNR}}{5}\bigg)^{-1.4}
\end{equation}
When matching GISMO-COSMOS sources with source catalogs from the literature, these positional uncertainties should be added in quadrature with the 1$\,\sigma_{\rm p}$ catalog position errors.

\subsection{Associations with known COSMOS (sub)mm sources} 
\label{sec:counterp}
\begin{figure*}
	\begin{center}
		\includegraphics[width=0.9\linewidth]{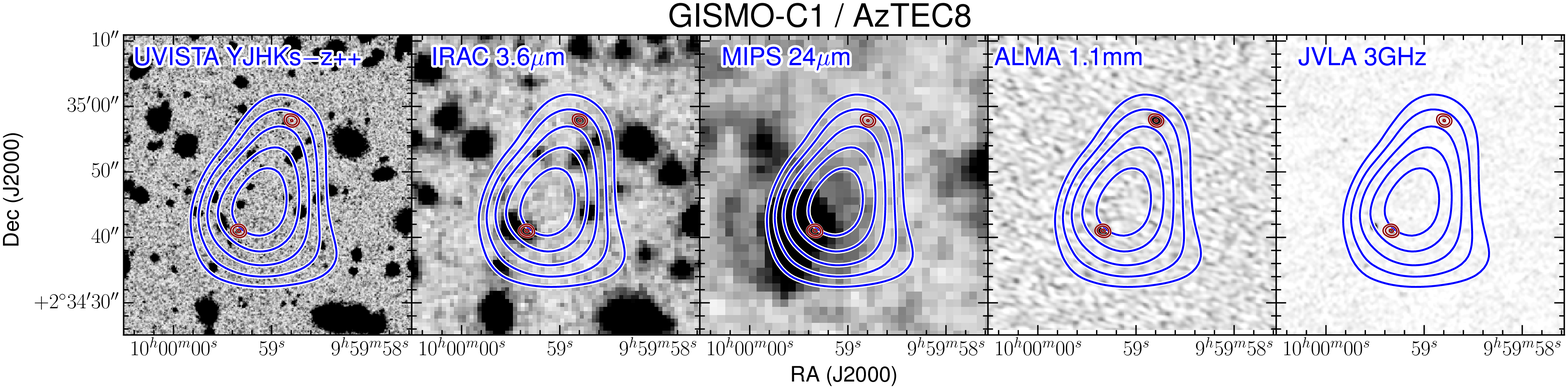}\vspace{0.1cm}        
		\includegraphics[width=0.9\linewidth]{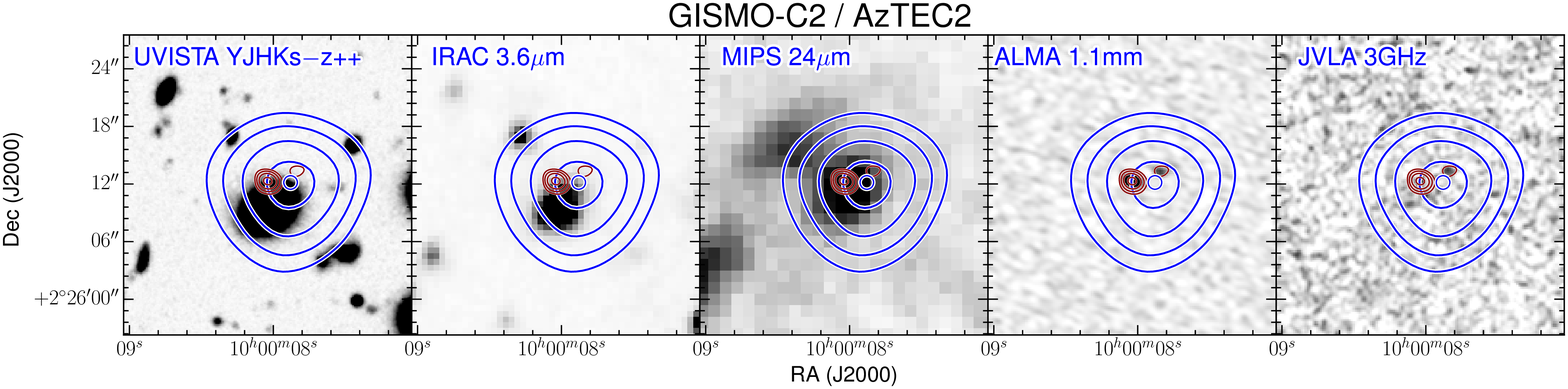}\vspace{0.1cm}        
		\includegraphics[width=0.9\linewidth]{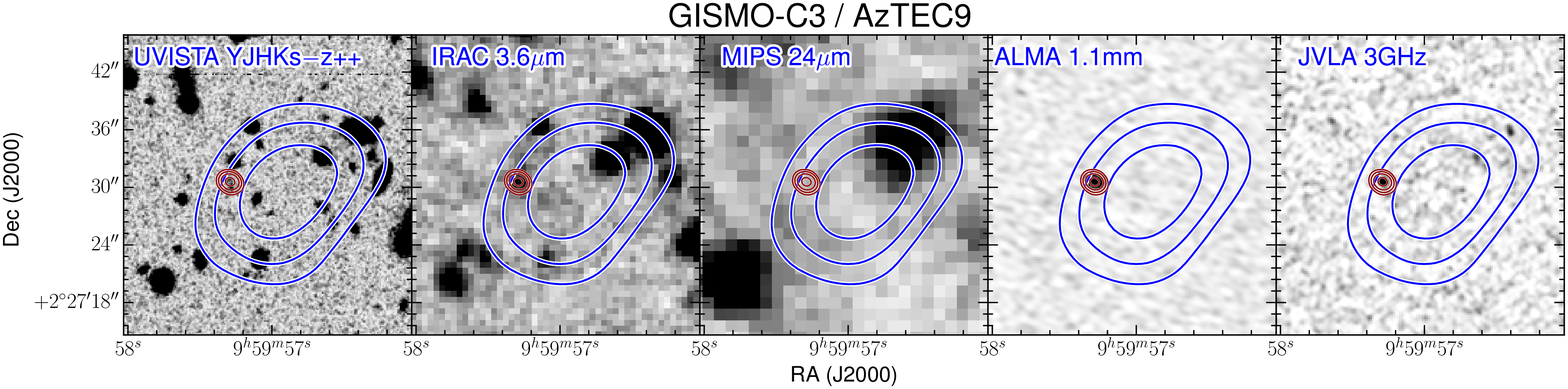}\vspace{0.1cm}        
		\includegraphics[width=0.9\linewidth]{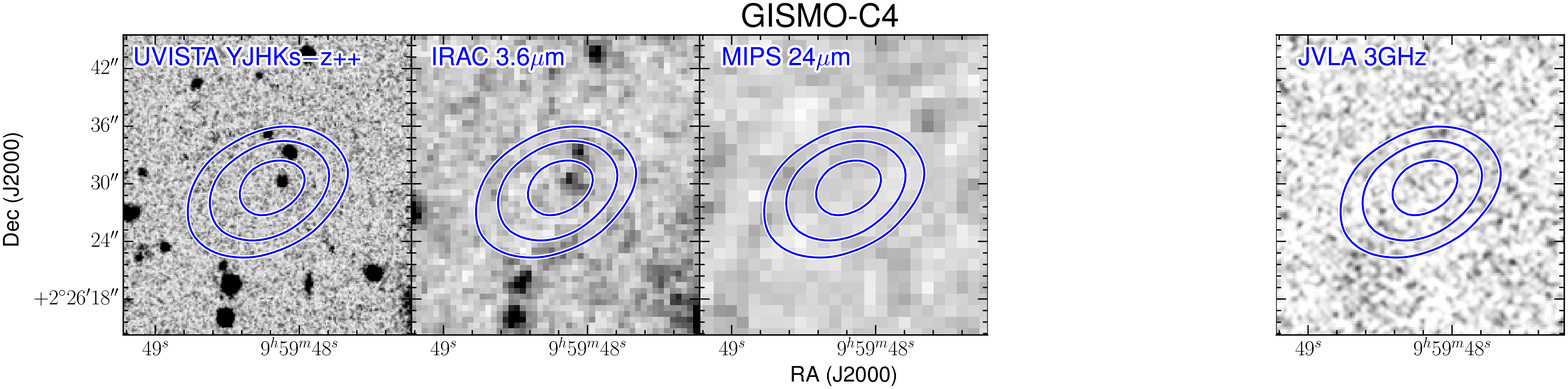}\vspace{0.1cm}        
		\includegraphics[width=0.9\linewidth]{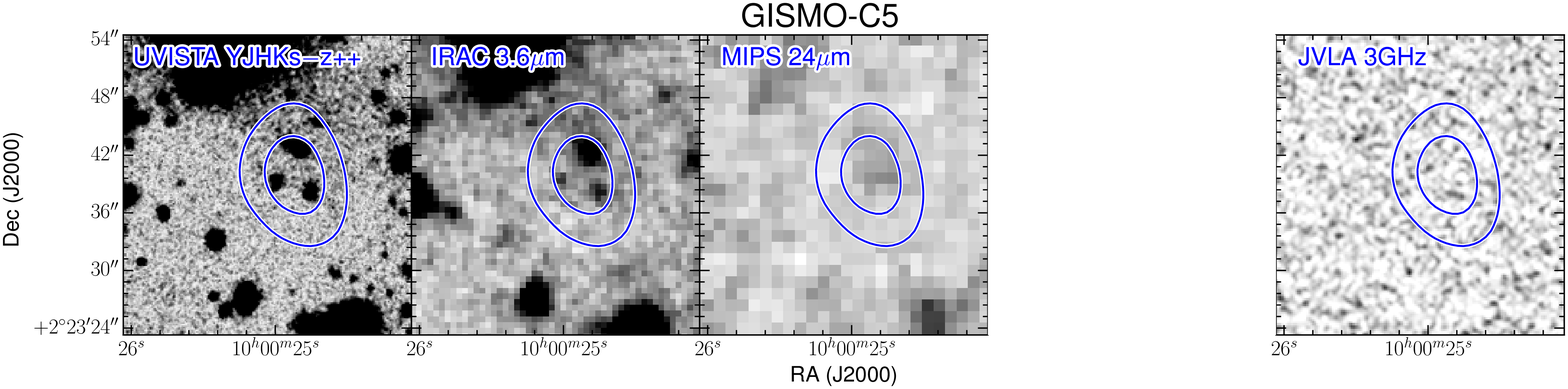}
        \caption{\label{fig:counterparts}
        		40\arcsec$\,\times$40\arcsec\ cutouts for $\geq\,$3.7$\sigma$ GISMO sources in the near-infrared (UVISTA YJHKs-z++), \textit{Spitzer}-IRAC 3.6$\,\mu$m, \textit{Spitzer}-MIPS 24$\,\mu$m, ALMA 1.3\,mm (when available) and JVLA 3\,GHz.
                Blue contours represent the flux levels in the GISMO map in 0.5$\,\sigma$ steps, starting at 3$\,\sigma$.
                The detection significance of each GISMO sources can thus be directly read off these contours.
                Red contours show the flux levels in the ALMA 1.3\,mm maps (when available) in 0.5$\,\sigma$ steps, starting at 3$\,\sigma$.
        		}
	\end{center}
\end{figure*}

We searched for the counterparts of our GISMO sources in all relevant (sub)mm catalogs publicly available in the COSMOS field, i.e., the 1.1mm AzTEC/JCMT \citep{scott_2008} and AzTEC/ASTE \citep{aretxaga_2011} surveys, and the deep {\sc Scuba}-2/JCMT 450\,$\mu$m and 850\,$\mu$m surveys \citep{casey_2013,geach_2017}.
Counterparts were deemed robust if their separations with our sources were lower than the quadratic combination of the GISMO positional uncertainties and the 1-$\sigma$ position errors of the relevant catalog\footnote{The source density in the AzTEC and {\sc Scuba}-2 catalogs are low enough that at this radius the probability that a counterpart is a random association is lower than 5\%.}.
We found counterparts for five GISMO sources: GISMO-C1/AzTEC8, GISMO-C2/AzTEC2, GISMO-C3/AzTEC9, GISMO-C7/AzTEC5 and GISMO-C6/\ SCUBA-2\,m450.173/850.104.
Amongst those, four benefit from intermediate resolution ($\sim1\arcsec-2\arcsec$) (sub)mm interferometric follow-up of AzTEC sources with the SMA at 890\,$\mu$m \citep{younger_2007, younger_2009} and ALMA at 1.3mm \citep{brisbin_2017}.
This follow-up is key to obtaining robust multi-wavelength and subsequently redshift identification of these galaxies, otherwise impossible with the coarse resolution of single-dish (sub)mm observations (Fig.~\ref{fig:counterparts}).
However, despite this effort, reliable spectroscopic redshift estimates are not yet available for all our sources, mostly due to their faint UV/optical/near-infrared counterparts \citep{casey_2017}.
Nevertheless, based on optical/NIR/FIR/mm/radio photometric information, there is an emerging consensus in the literature about their high-redshift nature \citep{koprowski_2014, miettinen_2015, brisbin_2017}.
In the following, we summarize the current knowledge on the five GISMO galaxies with (sub)mm counterparts and discuss the possible nature of the remaining four sources with no counterparts.
\begin{figure*}
\begin{center}
		\includegraphics[width=0.9\linewidth]{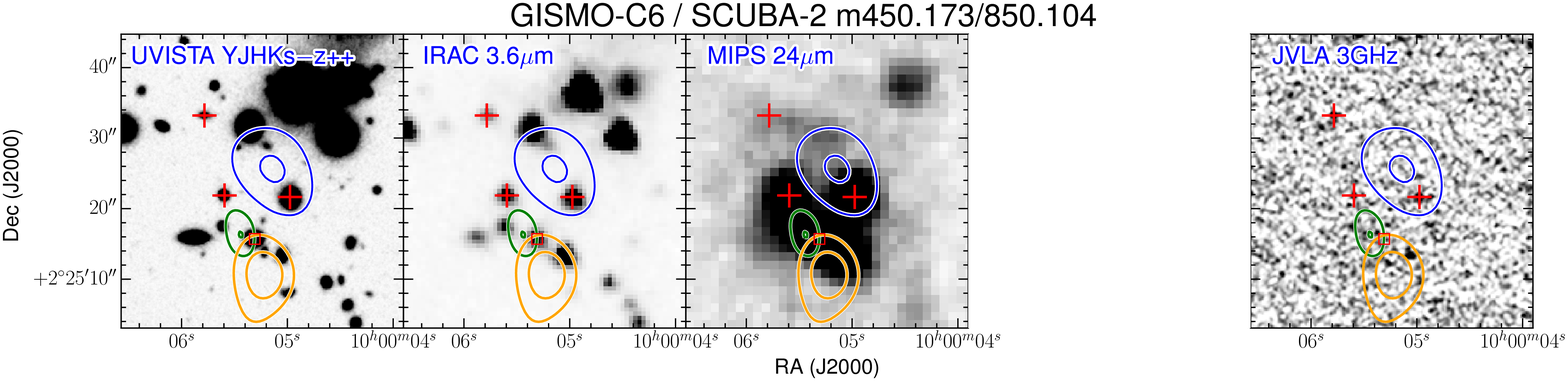}\vspace{0.1cm}
		\includegraphics[width=0.9\linewidth]{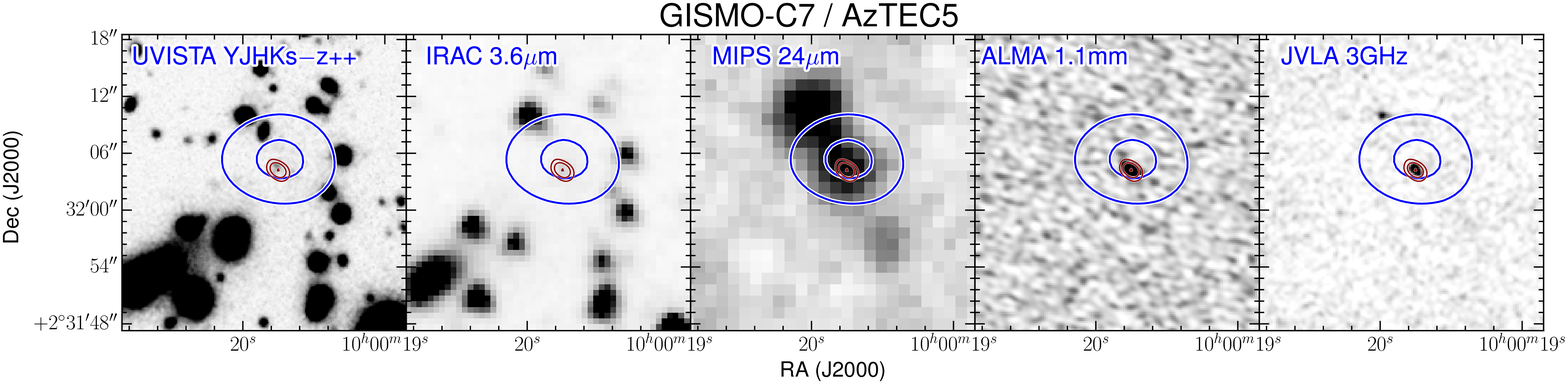}\vspace{0.1cm}
		\includegraphics[width=0.9\linewidth]{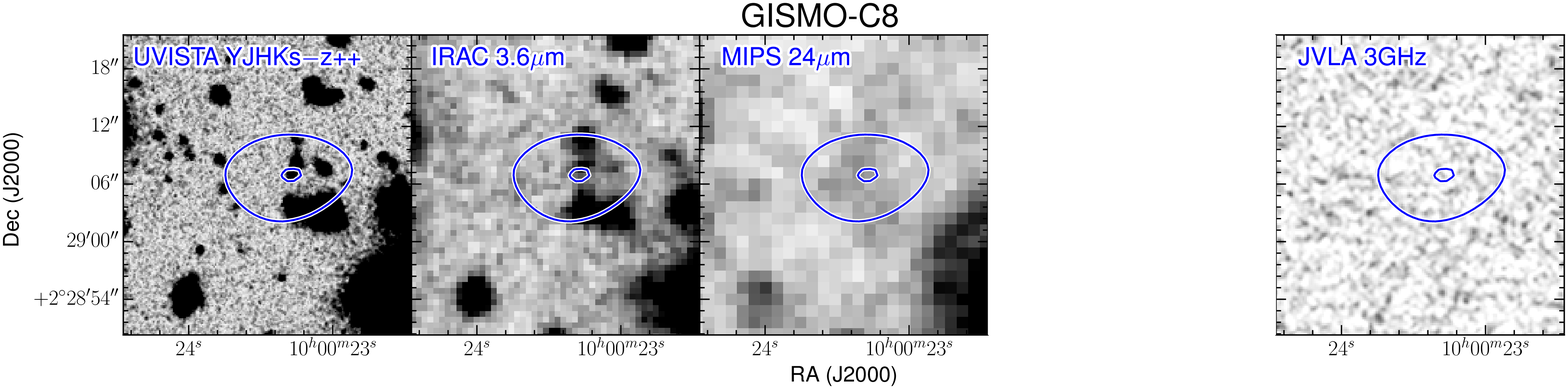}\vspace{0.1cm}
		\includegraphics[width=0.9\linewidth]{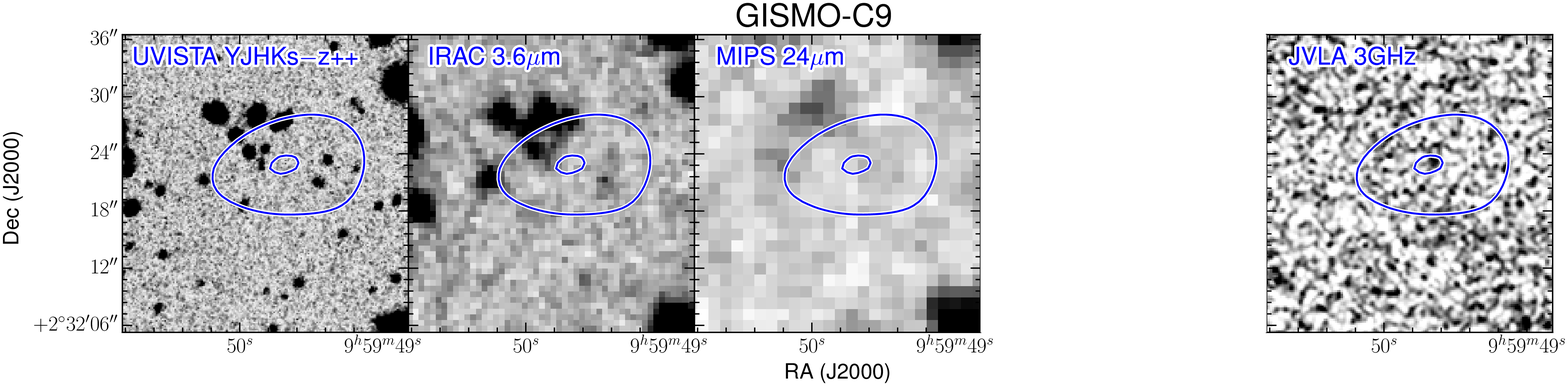}\\
\end{center}
		\textbf{Figure~\ref{fig:counterparts}} -- {\it continued.} -- For GISMO-C6, we show the optical counterpart of SCUBA-2\,m450.173/850.104 (red square) identified in \citet{casey_2013,casey_2017}, along with the JVLA-3\,GHz $\geq5$-$\sigma$ detections (red crosses). Finally, for this galaxy, we also add the 2.5$\sigma$ and 3$\sigma$ contours of the {\sc Scuba}-2 450$\,\mu$m (green) and 850$\,\mu$m (orange) maps from \citet{casey_2013}. 

\end{figure*}

\subsubsection{GISMO-C1\,/\,AzTEC8}
The brightest galaxy in our GISMO 2\,mm survey corresponds to the eighth and second brightest galaxies in the 1.1mm AzTEC/JCMT \citep[AzTEC8 -- $\theta_{\rm offset}$$\,=\,$$8\farcs4$;][]{scott_2008} and AzTEC/ASTE \citep[AzTEC-C2][]{aretxaga_2011} surveys, respectively.
It thus benefited from extensive (sub)mm interferometric follow-up \citep[e.g.,][]{younger_2007, younger_2009,brisbin_2017}, the latest being performed with ALMA at 1.3mm \citep{brisbin_2017,miettinen_2017c}.
It revealed two millimeter counterparts, with southern and northern component separated by $\sim\,$18\arcsec\ (Fig.~\ref{fig:counterparts}).
\cite{smolcic_2012} and \cite{koprowski_2014} reported a photometric redshift of $3.17^{+0.29}_{-0.22}$ and $3.15^{+0.05}_{-0.15}$ for the southern component, which in combination with a single CO line detection with CARMA leads to a best available solution of $z=3.179$ \citep[][Riechers et al. in prep]{smolcic_2012, brisbin_2017}.
The northern component is fainter at 1.3mm than the southern component, and has only a very uncertain photometric redshift estimate \citep[$z_{\rm phot}\lesssim3$;][]{brisbin_2017}.
This component fell below the AzTEC/JCMT detection threshold \citep{scott_2008} but contributed significantly to the flux density reported by \citet{aretxaga_2011} using the coarse resolution of the AzTEC/ASTE survey (34\arcsec\ vs. 17\arcsec).
Our GISMO-C1 detection also exhibits a slight extention towards the northern component.
Applying a PSF-fitting analysis to our GISMO map at the position of these two components, we found a very similar 2\,mm flux density for the southern component as that reported in Tab.~\ref{tab:sample} (1.33$\pm$0.22\,mJy vs. 1.29$\pm$0.22\,mJy), while the north component falls below our detection threshold, i.e., 0.65$\pm$0.23\,mJy.
Fitting only the southern component did not change its 2\,mm flux density, while fitting only the north component yields a $<\,$3.7$\,\sigma$ detection with bad residuals.
We thus conclude that the northern component did not significantly contribute to the 2\,mm flux density of GISMO-C1.
GISMO-C1 and AzTEC8 are most likely colocated near the southern component at a redshift of 3.179. 

As described in \citet{younger_2007,younger_2009}, about $\sim\,$2\arcsec\ east from GISMO-C1/AzTEC8 lies a bright and compact radio galaxy (Fig.~\ref{fig:counterparts}).
This galaxy is not associated with any millimeter emission but to a 24$\,\mu$m-bright low-redshift counterpart.
\textit{Spitzer} and \textit{Herschel} observations ($\sim\,$5--36\arcsec) are likely dominated by emission from this low-redshift interloper, and it is thus impossible to measure reliable 24-to-500$\,\mu$m flux densities for GISMO-C1/AzTEC8.
Using \textit{Spitzer} and \textit{Herschel} measurements as upper limits, we derived the dust mass and infrared luminosity of GISMO-C1/AzTEC8 via FIR-to-mm SED fitting using the dust model of \cite{draine_2007} (Tab.~\ref{tab:flux infrared}; Fig.~\ref{fig:sed}; see Sect.~\ref{subsec:SED} for details).
We found $\log(M_{\rm dust}/{\rm M_\odot})=9.9\pm0.1$ and $\log(L_{\rm IR}/{\rm L_\odot})=12.6\pm0.3$, yielding a ${\rm SFR=400^{+390}_{-200}\,{\rm M_\odot\,yr^{-1}}}$.
Finally, high-resolution ($<0\farcs05$) observations at $870\mu{\rm m}$ with ALMA have revealed asymmetric structures with multiple clumps in the central kilo-parsec of AzTEC8 \citep{iono_2016}. 

\subsubsection{GISMO-C2\,/\,AzTEC2/\,SCUBA-2\,450.03/850.00}
GISMO-C2 is associated with the second and third brightest galaxies in the 1.1mm AzTEC/JCMT \citep[AzTEC2 -- $\theta_{\rm offset}$$\,=\,$$2\farcs3$;][]{scott_2008,younger_2007} and AzTEC/ASTE \citep[AzTEC-C3;][]{aretxaga_2011} surveys, respectively. The source is also coincident with 450.03/850.00 in the SCUBA-2 450$\mu$m and 850$\mu$m maps of \citet{casey_2013}.
Follow-up with ALMA at 1.3mm revealed two components within the contours of our GISMO detection \citep[Fig.~\ref{fig:counterparts};][]{brisbin_2017}.
These two components are separated by 3\arcsec, the eastern component being $\sim\,$$\times\,$$4$ brighter at 1.3mm than the western component.
Both have counterparts in the 0\farcs75 JVLA 3\,GHz COSMOS survey \citep{smolcic_2017}.
Unfortunately, there are no reliable spectroscopic redshift estimates of either component. 
A preliminary optical/near-infrared redshift solution of $z=1.123$ has been used in the literature \citep[e.g.,][]{smolcic_2012,smolcic_2017, miettinen_2015,miettinen_2017,miettinen_2017b} for the eastern component; however, 
this solution corresponds to an optical counterpart 1\arcsec\ offset to the south of the eastern component.  That galaxy, visible in the optical behind what appears to be a much lower redshift ($z=0.3$) galaxy whose centroid is another 1\arcsec\ to the south, could have an associated tentative CO line emission detected with CARMA \citep[see discussion of the ambiguity of this source in][E.F. Jim\'enez-Andrade priv. comm.]{casey_2017}.  However, further analysis of this source's obscured SED argues against a $z=1.123$ redshift solution, given the unusually cold dust temperature that such a redshift would imply.
Without direct optical/near-infrared counterparts, the only redshift constraint we can place on GISMO-C2 are based on the ALMA-1.3\,mm and JVLA-3\,GHz radio photometry.
For the eastern component, \cite{brisbin_2017} found $z=3.89^{+3.11}_{-0.67}$, while for the western component they reported $z=2.03^{+1.19}_{-0.31}$.
Assuming that flux density ratios of these components are the same at 1.3mm and 2mm (i.e., 4.5/1.15), we concluded that GISMO-C2 is dominated by emission from the eastern component, i.e., a galaxy potentially at $z>3$.

It is clear that \textit{Spitzer} and \textit{Herschel} observations at the position of GISMO-C2/AzTEC2 are significantly contaminated by emission from the foreground galaxies, potentially both the $z=1.12$ system and the $z\sim0.3$ galaxy further to the south (Fig.~\ref{fig:counterparts}). 
Using \textit{Spitzer} and \textit{Herschel} measurements as upper limits, scaling all single-dish (sub)mm flux densities by the flux density ratio observed at 1.3mm, and using $z=3.89$, we found $\log(M_{\rm dust}/{\rm M_\odot})=9.6\pm0.1$ and $\log(L_{\rm IR}/{\rm L_\odot})=13.0\pm0.3$, that yields a ${\rm SFR=1000^{+1000}_{-500}\,{\rm M_\odot\,yr^{-1}}}$ (Tab.~\ref{tab:flux infrared}; Fig.~\ref{fig:sed}) for GISMO-C2/AzTEC2.

\subsubsection{GISMO-C3\,/\,AzTEC9/\,SCUBA-2\,850.01}
GISMO-C3 is associated with AzTEC9 \citep[$\theta_{\rm offset}$$\,=\,$$7\farcs1$;][]{scott_2008,younger_2009}, AzTEC/C14 \citep[][]{aretxaga_2011}, and SCUBA-2\,850.01 \citep{casey_2013}.
ALMA follow-up at 1.3mm revealed a single component, well within our GISMO contours \citep[Fig.~\ref{fig:counterparts};][]{brisbin_2017}.
A tentative spectroscopic redshift of 1.357 was obtained for AzTEC9 from a relatively weak spectrum (Salvato et al., in prep) with DEIMOS at the Keck Telescope.  
Although this estimate is consistent with the photometric redshift derived by \cite{smolcic_2012} of $1.07^{+0.11}_{-0.10}$, and the photometric redshift of $z=1.45$ from \citet{laigle_2016}, it has been suggested that this spectroscopic redshift corresponds to a nearby source unassociated with the submm emission \citep{koprowski_2014}.  It has also not been verified in deep near-infrared observations with Keck/MOSFIRE where one might have expected to cleanly detect H$\alpha$ emission \citep{casey_2017}. 
In contrast, \cite{brisbin_2017} have derived optical/NIR and FIR based photometric redshifts of $4.58^{+0.25}_{-0.68}$ and  $4.39\pm1.39$, respectively, which agree with those derived by \cite{koprowski_2014} of $4.85^{+0.50}_{-0.15}$ and $4.60^{+0.50}_{-0.31}$, respectively.
Based on the inconsistency of the low-redshift solution, we suspect the high-redshift solution for GISMO-C3/AzTEC9 is more likely.

GISMO-C3/AzTEC9 does not have any 24$\,\mu$m counterpart.
It is associated with emission in the \textit{Herschel} images but its non-detection in the {\sc Scuba}-2/JCMT 450\,$\mu$m image suggests that a significant fraction of these \textit{Herschel} flux densities comes from nearby galaxies.
Assuming $z=4.58$ and treating \textit{Herschel} flux densities as upper limits, we found $\log(M_{\rm dust}/{\rm M_\odot})=9.6\pm0.1$ and $\log(L_{\rm IR}/{\rm L_\odot})=13.0\pm0.2$, that yields a ${\rm SFR=1000^{+580}_{-370}\,{\rm M_\odot\,yr^{-1}}}$ (Tab.~\ref{tab:flux infrared}; Fig.~\ref{fig:sed}) for GISMO-C3/AzTEC9.

\subsubsection{GISMO-C6\,/\,SCUBA-2\,m450.173/850.104}
GISMO-C6 could be associated with SCUBA-2 m450.173/850.104 ($\theta_{\rm offset}$$\,=\,$$7\farcs5$), one of the marginal 3$\,<\,$$\sigma$$\,<\,$3.6 450$\,\mu$m identified sources with $>\,$3$\sigma$ 850$\,\mu$m counterparts reported by \citet{casey_2013}.
Though there are only marginal detections at all of these wavelengths, the source is unlikely to be spurious for having been detected in multiple independent datasets.
The source is ambiguous because GISMO-C6 is situated $\sim\,$7\farcs5 away from the SCUBA-2 position, though this is within the positional uncertainties of both sources.
\citet{casey_2013} associated SCUBA-2\,m450.173/850.104 to an optical counterpart with 24$\,\mu$m counterpart and $z_{\rm phot}=1.01$ (Fig.~\ref{fig:counterparts}).
Near-infrared spectroscopic follow-up of this source with Keck MOSFIRE in \citet{casey_2017} yields a spectroscopic redshift of 1.003 and subsequently a infrared luminosity of $\log(L_{\rm IR}/{\rm L_\odot})=11.81\pm0.25$.
However, deep radio 3\,GHz imaging of the COSMOS field \citep{smolcic_2017}, did not show any radio counterpart to this source, leading to a far-infrared-to-radio luminosity ratio, $q$, of $2.8$ at odds with the expected $2.47\pm0.26$ value at this redshift \citep{magnelli_2015,delhaize_2017}.
The 3\,GHz COSMOS image shows, however, two other possible counterparts (Fig.~\ref{fig:counterparts}) with a photometric redshift of $0.78$ and $1.00$, $\sim\,$6\arcsec\ northwest and $\sim\,$6\arcsec\ northeast from the source identified by \citet{casey_2017}, respectively.
\textit{Spitzer} and \textit{Herschel} observations exhibit bright but confused emission most likely associated with these low-redshift galaxies.
Similarly, the (sub)mm emission of GISMO-C6/SCUBA-2\,m450.173/850.104 could well originate from the combined emission of these low-redshift galaxies, which might be part of a single system. 
However, it could also be emitted by a yet unknown optically-faint high-redshift galaxy.
Deep interferometric (sub)mm observations are needed to further investigate the nature of GISMO-C6\,/\,SCUBA-2\,m450.173/850.104. 

Assuming a low redshift origin of the (sub)mm emission, i.e., $z=1.003$ as in \citet{casey_2017}, and using \textit{Herschel} measurements as upper limits, we found $\log(M_{\rm dust}/{\rm M_\odot})=9.2\pm0.3$ and $\log(L_{\rm IR}/{\rm L_\odot})=11.6\pm0.3$, that yields a ${\rm SFR=40^{+40}_{-20}\,{\rm M_\odot\,yr^{-1}}}$ (Tab.~\ref{tab:flux infrared}; Fig.~\ref{fig:sed}).
These estimates are fully consistent with that inferred in \citet{casey_2017}.

\subsubsection{GISMO-C7\,/\,AzTEC5/\,SCUBA-2\,450.04/850.03}
GISMO-C7 is associated with AzTEC5 \citep[$\theta_{\rm offset}$$\,=\,$$1\farcs6$;][]{scott_2008}, AzTEC/C42 \citep[][]{aretxaga_2011}, and SCUBA-2\,450.04/850.03 \citep{casey_2013}.
The source has interferometric follow-up from both the SMA at 890$\,\mu$m \citep{younger_2007,younger_2009} and from ALMA at 1.3mm \citep[Fig.~\ref{fig:counterparts};][]{brisbin_2017} revealing a single counterpart.
Although there still is no spectroscopic redshift reported in the literature for AzTEC5, \citet{casey_2013} quote a optical/near-infrared photometric redshift, derived in \citet{ilbert_2013}, of $z_{\rm phot}=3.82^{+0.44}_{-0.69}$, which is consistent with other far-infrared/radio determinations of the photometric redshift \citep{smolcic_2012, koprowski_2014, brisbin_2017}.
Recently, thanks to high-resolution observations with HST towards AzTEC5, \citet{gomez_2018} reported that it is has three primary rest-frame UV components, with stellar masses of $\log(M_\star/M_\odot)=9.92^{+0.10}_{-0.10},\,9.78^{+0.08}_{-0.10}$ and $9.59^{+0.08}_{-0.06}$ and 3D-HST-based redshifts of $z = 3.63^{+0.14}_{-0.15}$,  $z = 4.02^{+0.08}_{-0.08}$ and  $z = 3.66^{+0.40}_{-0.43}$ \citep{brammer_2012, skelton_2014,moncheva_2016}. 
Since these redshifts are consistent within the uncertainties, it is likely that these three components belong to the same system \citep{gomez_2018}, favoring a scenario in which GISMO-C7/AzTEC5 is a merger-driven star-forming galaxy at $z\sim3.6$.

Assuming $z=3.63$ and treating \textit{Herschel} measurements as upper limits, we found $\log(M_{\rm dust}/{\rm M_\odot})=9.3\pm0.1$  and $\log(L_{\rm IR}/{\rm L_\odot})=13.1\pm0.1$, that yields a ${\rm SFR=1250^{+320}_{-260}\,{\rm M_\odot\,yr^{-1}}}$ (Tab.~\ref{tab:flux infrared}; Fig.~\ref{fig:sed}) for GISMO-C7/AzTEC5.\\

\subsubsection{GISMO sources with no (sub)mm counterparts}
Four out of nine GISMO $\geq\,$$3.7$$\sigma$ detections are not associated with any known SMGs, even though they are within deep and available (sub)mm coverage of the COSMOS field \citep{scott_2008, aretxaga_2011, casey_2013,geach_2017}.
Such (sub)mm dropouts could be false 2\,mm detections.
However, the probability of all four being false detections is relatively low, as only 9\% of our pure jackknifed maps have $\geq$4 false detections with S/N$\,\geq3.7$. 
These (sub)mm dropouts could instead be unidentified high-redshift galaxies (e.g., S14).

GISMO-C4 is the fourth brightest source in our survey.
It is detected with high significance and is thus very unlikely to be a false detection, i.e., $P_{\rm f}$$\,=\,$$6.2$\% (Tab.~\ref{tab:sample}).
It is within deep {\sc Scuba}-2/JCMT and AzTEC/ASTE coverages \citep[i.e., $\sim 0.8$ and $\sim 1.26\,$mJy/beam rms at 850$\,\mu$m and 1.1mm, respectively;][]{casey_2013,aretxaga_2011}.
Taking the 80\% completeness detection limits of these surveys\footnote{these 80\% completeness limits are read off Fig.~6 of \citet{casey_2013} -- taking the mean value from their 3$\,\sigma$ and 4$\,\sigma$ curves as the final 850\,$\mu$m catalog is cut at 3.6$\,\sigma$ -- and Fig.~5 of \citet{aretxaga_2011}} (i.e., 5 and 5.5\,mJy at 850$\,\mu$m and 1.1\,mm, respectively), yields 850$\,\mu$m-to-2\,mm and 1.1\,mm-to-2\,mm flux densities ratios $<\,$7.1 and $<\,$7.8, respectively. 
Such low flux densities ratios indicate that the 850$\,\mu$m and 1.1\,mm broadbands do not probe the Rayleigh-Jeans emission of GISMO-C4, because otherwise the 850$\,\mu$m-to-2\,mm and 1.1\,mm-to-2\,mm flux densities ratios would be in the range 20--30 and 8 -- 11, respectively.
Instead, the 850$\,\mu$m and 1.1\,mm broadbands likely probe rest-frame wavelengths shortward of 300$\,\mu$m, supporting a high-redshift solution for this galaxy, i.e., $z>3-4$.
This high-redshift candidate will require dedicated follow-up efforts with, e.g., ALMA or NOEMA.

GISMO-C5, GISMO-C8 and GISMO-C9 are all detected at 4.0$\,\geq\,$S/N$\,\geq\,$3.7 at 2\,mm, a detection significance range in which we expect 1.22 false detections (Sect.~\ref{subsubsec:contanimation}).  
Given that GISMO-C6 and GISMO-C7 (the other two sources detected in this significance range) have known (sub)mm counterparts, it is likely that at least one of these three (sub)mm dropouts is a false detection.
GISMO-C5, GISMO-C8 and GISMO-C9 are within deep AzTEC/ASTE coverage \citep[][]{aretxaga_2011}.
However, only GISMO-C5 and GISMO-C8 sit in the deep central 850$\mu$m map of \citet{casey_2013}, while GISMO-C9 has shallower 850$\mu$m upper limit from \citet{geach_2017}.
The 80\% completeness detection limits from these surveys\footnote{the 80\% completeness from \citet{geach_2017} is read off their Fig.~8, i.e., 6.2\,mJy}, lead to 850$\,\mu$m-to-2\,mm and 1.1\,mm-to-2\,mm flux densities ratios in the range 9.8--13.3 and 10.7--11.8, respectively.
Such low flux densities (yet not as low as observed in GISMO-C4) suggests that these galaxies reside at high redshift (i.e., $z\gtrsim3$).
Further follow-up efforts are needed, keeping in mind that at least one of these sources is likely a false detection.

\section{Results}
\label{sec:results}

\subsection{2\,mm source counts}
\label{subsec:counts}
In the absence of robust redshift determinations for all our sources, number counts are the most powerful tool to constrain models of galaxy evolution from our catalog.
``Corrected'' cumulative number counts, i.e., $N(>$$\,S)$, are given by the number of galaxies with a flux density higher than $S$:
\begin{equation}
N(>S) = \sum^{S_{\rm in}^{i}>S}{\frac{ 1-P_{f}(S^{i}_{\rm out}/N^{i}_{\rm out}) }{A_{\rm eff} \times C(S^{i}_{\rm in})}},
\end{equation}
where $S_{\rm in}^{i}$ and $S^{i}_{\rm out}$ are respectively, the \textit{deboosted} and \textit{observed} flux densities of the $i$th source, while $N^{i}_{\rm out}$ is its \textit{observed} flux densities error  (Tab.~\ref{tab:sample}); $P_{f}(S^{i}_{\rm out}/N^{i}_{\rm out})$ is its probability to be a false detection (Tab.~\ref{tab:sample}); $C(S^{i}_{\rm in})$ is the completeness of our source extraction at this deboosted flux density (Fig.~\ref{fig:Completeness}); and, finally, $A_{\rm eff}$ is the area for the source extraction (i.e., 250\,arcmin$^2$ where the rms is better than 0.23 mJy/beam).
Deboosted flux densities and $P_{f}(S^{i}_{\rm out}/N^{i}_{\rm out})$ being model-dependent, ``corrected'' cumulative number counts must be evaluated for each model.
To avoid these model-dependencies, we will also report here ``raw'' number counts, by setting $P_{f}(S^{i}_{\rm out}/N^{i}_{\rm out})=0.0$, $C(S^{i}_{\rm in})=1$ and $S_{\rm in}^{i}=S^{i}_{\rm out}$. 
``Raw'' number counts are not easily comparable with past and future literature measurements as these latter naturally suffer from different observational biases (i.e., flux boosting, contamination and completeness).
However, ``raw'' number counts can be compared to model predictions using the mock maps produced in Sect.~\ref{subsec:extraction} and which reproduce the observational biases affecting our survey. 

\begin{figure*}
\begin{center}
\includegraphics[width=0.85\linewidth]{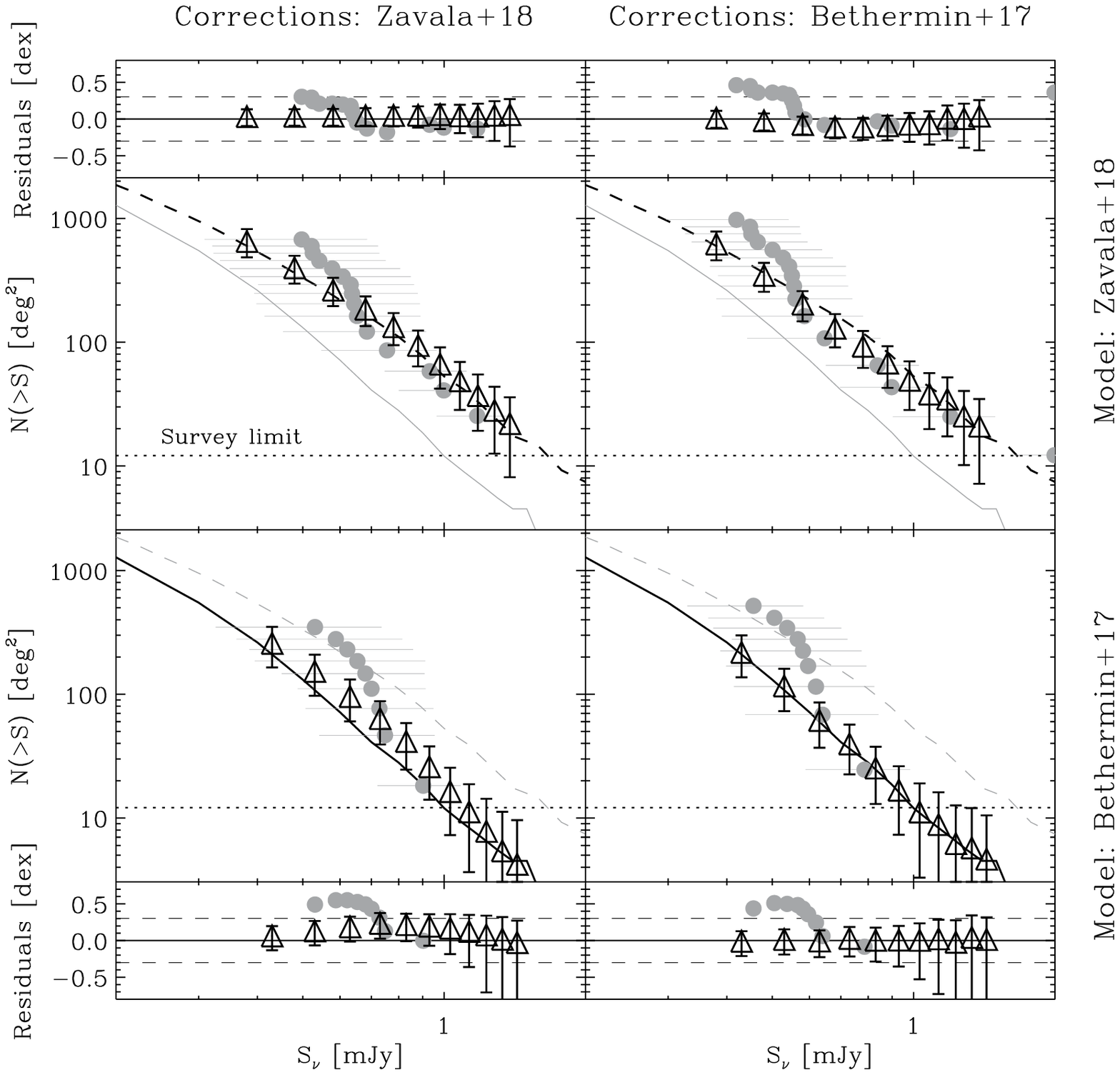}		
\caption{\label{fig:test method}
		``Corrected'' cumulative number counts (empty triangles) measured on \textbf{simulated maps} following the models of \citet{zavala_2018} (upper row) and \citet{bethermin_2017} (lower row), using our Monte-Carlo approach and corrections inferred from the models of \citet{zavala_2018} (left column) and \citet{bethermin_2017} (right column).
        The input cumulative number counts followed by the models of \citet{zavala_2018} and \citet{bethermin_2017} are shown by the dashed and solid lines, respectively.
        The horizontal dotted line shows the limit of our survey, i.e., the sky density below which the number of sources in our simulated maps would be lower than 1.
        The residual between the inferred and input number counts are displayed in the upper and lower row panels.
        Grey circles show the ``corrected'' cumulative number counts measured in one of these \textbf{simulated maps} using the methodology described in S14 instead of our Monte-Carlo approach.
        With our Monte-Carlo methodology, the ``corrected'' number counts are consistent within the uncertainties with the input number count distribution.
        }
\end{center}
\end{figure*}

To properly account for the large, asymmetric and non-Gaussian deboosted flux uncertainties of our sources, we measured $N(>S)$ using a Monte-Carlo approach.
We created 1,000 realizations of our catalog, drawing the deboosted flux densities of each source following the flux boosting distribution measured at their respective S/N and shown in Fig.~\ref{fig:sout_sin}.
The ``corrected'' cumulative number counts and associated uncertainties are then given by the average and dispersion of the $N(>S)$ distribution measured over these 1,000 realizations.
To validate this approach, we applied this methodology to our simulations.
In the upper-left and lower-right panels of Fig.~\ref{fig:test method}, we show the mean and dispersion of $N(>S)$ as measured by applying this methodology to 200 map realizations.
On average, this methodology retrieved perfectly the input number count distribution and for 68\% of our map realizations it provides measurements within $\sim\,$0.15\,dex.
However, these tests correspond to an ideal case in which the corrections (i.e., flux boosting statistics and $P_{f}(S^{i}_{\rm out}/N^{i}_{\rm out}$)) were measured on the same model.
In reality, we obviously do not know the intrinsic sky model and are thus limited by the model-dependencies of our corrections.
To test this effect, we applied corrections measured on one model to the other model.
Results are shown in the upper-right and lower-left panels of Fig.~\ref{fig:test method}.
In these more realistic cases, we naturally do not retrieve perfectly the input number count distributions.
The corrections from \citet{zavala_2018}, with their lower flux boosting statistics, slightly over-estimate the number count distribution when applied to catalogs extracted from \citet{bethermin_2017} simulated maps.
On the other hand, corrections from \citet{bethermin_2017} slightly under-estimate the number counts when applied to the catalogs extracted from \citet{zavala_2018} simulated maps.
These over/under-estimations are, however, comparable to the uncertainties, and certainly, do not lead to number count distributions equal to that from the correction model.
Note that the number count distributions followed by these models bracket that inferred from our real catalog, suggesting that they provide a realistic representation of the range of possible corrections.

Finally, using the same simulations, we tested the methodology advocated in S14, i.e., plotting the number of sources at each deboosted flux density, divided by the effective area for the detection of sources.
Results of this test for one of our simulations are shown by grey circles in Fig.~\ref{fig:test method}.
Irrespective of the corrections used in this methodology, it systematically overestimates the number counts at low flux densities, where the uncertainties on the deboosted fluxes are large.
This is understandable when considering the case of two sources with the same deboosted flux densities but large uncertainties: at their common flux density, the number of sources is equal to two but the probability that both have fluxes greater than this value is equal to 0.5, leading to an $\times$2 over-estimation of the cumulative number counts.
Note that this test does not imply that the deboosted flux densities quoted in S14 are incorrect; simply that the number counts inferred in S14 from these deboosted flux densities are overestimated at faint flux densities.

Having validated our Monte-Carlo methodology on simulations, we now measure the ``raw'' and ``corrected'' GISMO COSMOS 2\,mm cumulative number counts using this approach.
For our ``raw'' measurements, this simply implies to create 1,000 realizations of our catalog, drawing the observed flux density of each source following Gaussian distributions characterized by their observed flux uncertainties. \\

\begin{figure*}
\centering
\includegraphics[width=\linewidth]{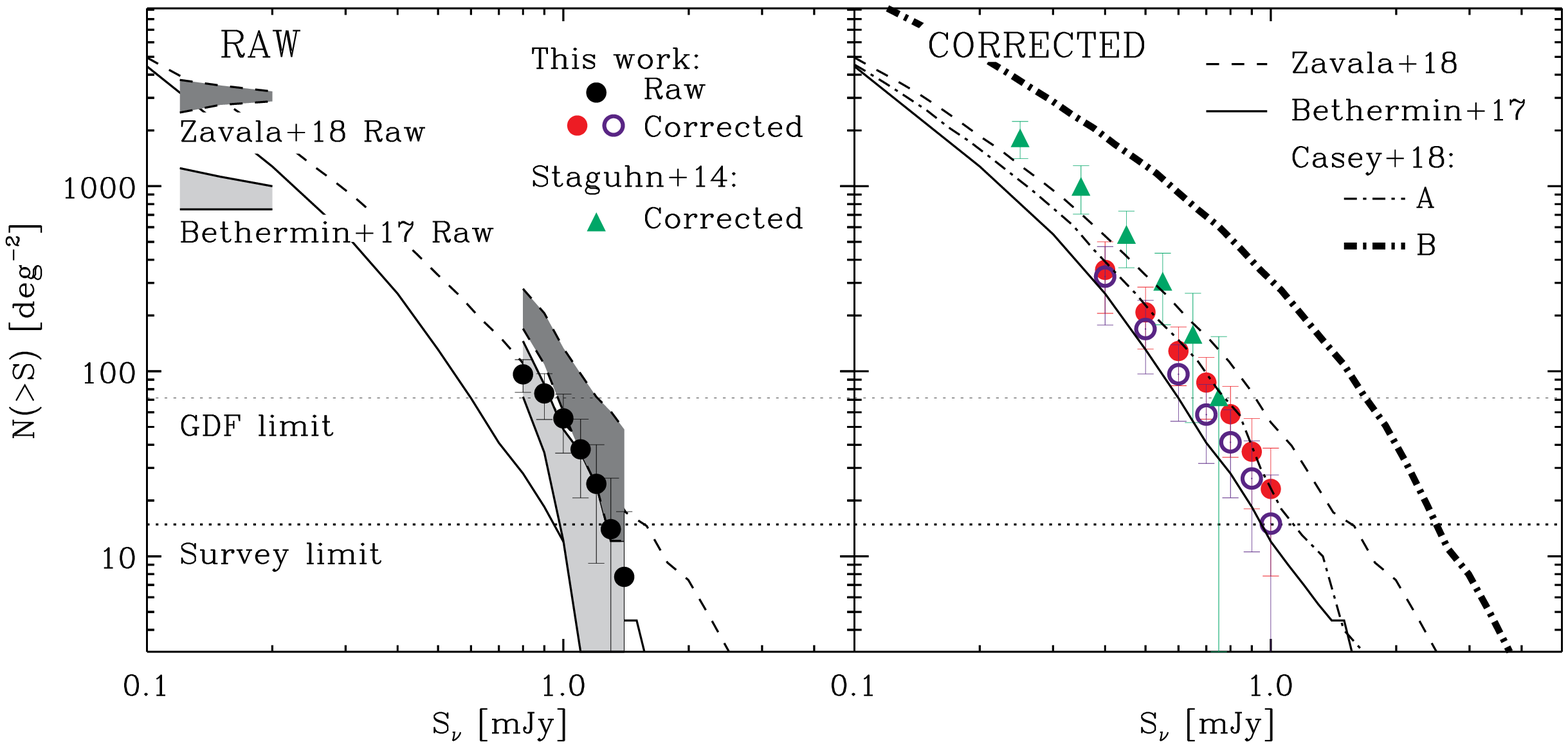}		
\caption{\label{fig:counts}
		Cumulative number counts measured by applying our Monte-Carlo methodology to the GISMO COSMOS catalog.
		In the left panel, black circles show the ``raw'' cumulative number counts, i.e., applying our Monte-Carlo methodology to our \textit{observed} flux densities \textit{without} completeness and contamination corrections.
        The dark-gray and light-gray regions show the 16th and 86th percentiles of the ``raw'' cumulative number count distributions measured using 1,000 mock catalogs from 1,000 simulated maps following the models of \citet{zavala_2018} and \citet{bethermin_2017}, respectively.
        In the right panel, red and opened blue circles present the ``corrected'' cumulative number counts measured by applying our Monte-Carlo methodology to our \textit{deboosted} flux densities and using completeness and contamination corrections from the models of \citet{zavala_2018} and \citet{bethermin_2017}, respectively. 
        Green triangles are data from S14, using their \citet{bethermin_2011}-based \textit{deboosted} fluxes.
        Cumulative number counts estimated by the models of \citet{zavala_2018} and \citet{bethermin_2017} are shown by the dashed and solid lines, respectively.
        In the right panel, the thin and thick dot-dashed lines show the cumulative number counts as in the model A and B of \cite{casey_2018}.
        The horizontal dotted line shows the limit of our survey, i.e., the sky density below which the number of sources in our map would be lower than 1.
        For comparison, we also show the survey limit of the GDF.
        ``Corrected'' cumulative number counts measured from our map using different corrections are consistent within the uncertainties with each other; as well as with the models of \citet{zavala_2018} and \citet{bethermin_2017} used to infer our corrections.
        Combined with the GDF measurements, we constrain the 2\,mm cumulative number counts over one decade in flux density.}
        \vspace{0.5cm}
\end{figure*}
\begin{table}
\begin{center} 
\caption{\label{tab:number counts} Raw and corrected cumulative number counts at 2\,mm }
\begin{tabular}{ c c c c c c} 
\hline \hline
\multicolumn{2}{c}{Raw number counts} & & \multicolumn{3}{c}{Corrected number counts} \\ [1ex]
\cline{1-2}
\cline{4-6}
\rule{0pt}{3ex} $S_{2\,{\rm mm}}^{\rm obs}$ & $N(>$$\,S)$ & & $S_{2\,{\rm mm}}^{\rm deboosted}$ & $N(>$$\,S)^{\rm a}$ & $N(>$$\,S)^{\rm b}$\\
\rule{0pt}{3ex}(mJy) & (deg$^{-2}$) & & (mJy) & (deg$^{-2}$) & (deg$^{-2}$) \\ [1ex]
\hline
\multicolumn{6}{c}{This Survey} \\ [1ex]
0.8 & 96$\,\pm\,$19 & & 0.4 & 353$\,\pm\,$147 & 324$\,\pm\,$147 \\[1ex]
0.9 & 76$\,\pm\,$21 & & 0.5 & 208$\,\pm\,$77  & 169$\,\pm\,$72 \\[1ex]
1.0 & 55$\,\pm\,$19 & & 0.6 & 129$\,\pm\,$45  & 96$\,\pm\,$43 \\[1ex]
1.1 & 38$\,\pm\,$17 & & 0.7 & 87$\,\pm\,$32   & 58$\,\pm\,$27 \\[1ex]
1.2 & 25$\,\pm\,$15 & & 0.8 & 59$\,\pm\,$24   & 41$\,\pm\,$21 \\[1ex]
1.3 & 14$\,\pm\,$13 & & 0.9 & 37$\,\pm\,$19   & 26$\,\pm\,$16 \\[1ex]
1.4 & $<\,$18       & & 1.0 & 23$\,\pm\,$15   & 15$\,\pm\,$12 \\ [1ex]
\hline
\multicolumn{6}{c}{GDF} \\ [1ex]
 &  & & 0.25 & 1823$\,\pm\,$415 \\ [1ex]
 &  & & 0.35 & 997$\,\pm\,$291 \\ [1ex]
 &  & & 0.45 & 548$\,\pm\,$185 \\ [1ex]
 &  & & 0.55 & 306$\,\pm\,$128 \\ [1ex]
 &  & & 0.65 & 158$\,\pm\,$105 \\ [1ex]
 &  & & 0.75 & 73$\,\pm\,$81 \\ [1ex]
\hline
\end{tabular}
\end{center}
\textbf{Notes.}
$^{(a)}$ Measured using corrections from the model of \citet{zavala_2018}, and for the GDF the model of \citet{bethermin_2011}.
$^{(b)}$ Measured using corrections from the model of \citet{bethermin_2017}.
\end{table}

In the left panel in Fig.~\ref{fig:counts} we show our ``raw'' 2\,mm cumulative number counts, while on the right panel we show our ``corrected'' cumulative number counts.
Table~\ref{tab:number counts} provides the same data in tabular form.
In the left panel of Fig.~\ref{fig:counts}, we compare these ``raw'' measurements with the ``raw'' predictions from the models of \citet{zavala_2018} and \citet{bethermin_2017}. 
These predictions correspond to the 16th and 86th percentiles of the ``raw'' cumulative number counts distribution measured using 1,000 mock source catalogs retrieved from 1,000 simulated maps generated in Sect.~\ref{subsec:extraction}.
On the right-hand panel, we compare our ``corrected'' 2\,mm cumulative number counts with the intrinsic model predictions.
Finally, on the right-hand panel we also compare our ``corrected'' measurements with those from S14, applying our Monte-Carlo methodology to their \citet{bethermin_2011}-based deboosted flux densities, assuming that their associated uncertainties follow a Gaussian distribution.

Our ``raw'' cumulative number counts are nicely bracketed by the ``raw'' predictions from the models of \citet{zavala_2018} and \citet{bethermin_2017}. 
This suggests that these models provide us with reasonable range of possible corrections, and thus robust ``corrected'' cumulative number count measurements.
This also suggests that models predicting significantly more or significantly fewer 2\,mm sources in the $\sim$mJy regime are inconsistent with our observations. 

Our two different sets of corrections yield very consistent ``corrected'' cumulative number count measurements, well within their uncertainties.
These measurements agree also within their uncertainties with those from S14 in the flux density range where they overlap.
This comparison clearly illustrates the advantage of the "wedding cake" observing strategy followed by the GISMO team.
On the one hand, the COSMOS map with its large sky coverage provides critical constraints on the 2\,mm number counts at high flux densities, inaccessible to the pencil-beam survey of the GDF.
On the other hand, the GDF provides critical constraints at low flux densities, well below the detection threshold of our COSMOS map.
Combining these two surveys, we obtain robust measurements of the 2\,mm number counts over almost one decade in flux density.

Both the models of \citet{zavala_2018} and \citet{bethermin_2017} are consistent within the uncertainties with our ``corrected'' measurements.
We also include two additional models drawn from \citet{casey_2018}, which are identical to the \citet{zavala_2018} model but assume a different evolution in $\Phi_\star$ of the obscured luminosity function beyond $z>2$: Model A represents a ``dust-poor'' early Universe, while Model B represents a ``dust-rich'' early Universe.
Among all models plotted in Fig.~\ref{fig:counts}, the \citet{casey_2018} Model A provides the best description of our estimates\footnote{Performing our Monte-Carlo simulations using the model A of \citet{casey_2018}, we ended up with deboosted flux densities and ``corrected'' number count measurements intermediate to those inferred from the models of \citet{zavala_2018} and \citet{bethermin_2017}, leaving our results unchanged}. 
Although with our number counts we cannot fully discriminate between the model of \citet{zavala_2018}, \citet{bethermin_2017} and Model A of \citet{casey_2018}, our data are inconsistent with the most extreme extrapolation of the \citet{casey_2018} Model B, the ``dust-rich'' early Universe.
It is important to point out, however, that such ``dust-rich'' models are very dependent on the model cutoff redshift \citep[discussed more extensively in][]{zavala_2018}.
This cutoff redshift represents a model instantaneous redshift above which no more DSFGs can be found, and letting it vary between $6<z_{\rm cutoff}<12$ changes the expected 2\,mm number counts of a ``dust-rich'' Universe substantially.
By tuning down this parameter, starting from its original value of 12 in \citet{casey_2018} and until we find a broad agreement with our number count measurements\footnote{The results of this fine tuning is, however, not shown in Fig.~\ref{fig:counts}.}, we can rule out Model B with $z_{\rm cutoff}\geq7$.

\begin{figure*}
\includegraphics[width=\linewidth]{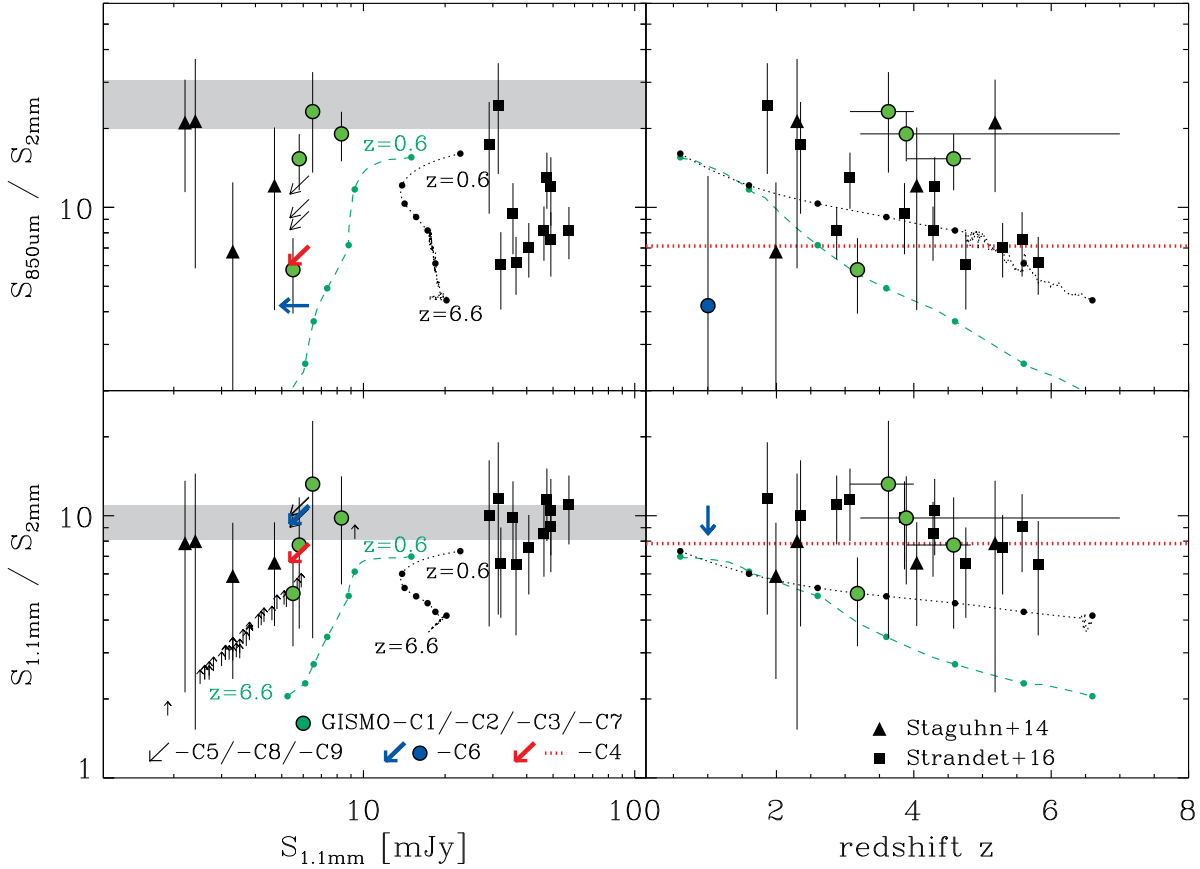}		
\caption{\label{fig:flux ratio}
		The 850$\,\mu$m-to-2\,mm (top row) and 1.1\,mm-to-2\,mm (bottom row) flux density ratios of (sub)mm-selected galaxies as a function their 1.1\,mm flux densities (left column) and redshifts (right column).
        Green filled circles show the GISMO COSMOS galaxies with AzTEC counterparts (i.e., GISMO-C1/-C2/-C3/-C7).
        GISMO-C6, which has a counterpart at 850$\,\mu$m but none at 1.1\,mm is shown by blue arrows and circle.
        GISMO-C4, our fourth brightest detection, which has no counterpart at 850$\,\mu$m and 1.1\,mm is shown by red arrows (left-hand panels) and red horizontal dotted lines (right-hand panels). 
        GISMO-C5/-C8/-C9 with no counterparts at 850$\,\mu$m and 1.1\,mm are displayed by black downward pointing arrows.
        Black triangles are from S14, while squares are from \citet{strandet_2016} extrapolating their 1.4\,mm into 1.1\,mm flux densities using $S_{\rm 1.1mm}=S_{1.4}\times(1.4/1.1)^{3.75}$.
        Black upward pointing arrows are 1.1\,mm sources in \citet{aretxaga_2011} not detected within our GISMO COSMO map.
        The gray shaded areas correspond to flux density ratio ranges expected if both broadbands would probe the Rayleigh-Jeans dust emission of galaxies with a dust emissivity of 1.5$\,\leq\,$$\beta$$\,\leq\,$2.0.
        The black dotted (green dashed) line shows the evolution with redshifts, from $z$$\,=\,$0.6 to $z$$\,=\,$6.6, of these flux density ratios for a galaxy having $L_{\rm IR}=10^{13.5}\,$L$_{\odot}$ ($10^{12.5}\,$L$_{\odot}$) and the same SED as Arp 220 \citep[Sd galaxy;][]{polletta_2007}.
        In the righ-hand panels, most galaxies have higher 850$\,\mu$m-to-2\,mm and 1.1\,mm-to-2\,mm flux density ratios than predicted from these SED templates, suggesting hotter dust temperatures at high-redshift than in the local Universe \citep[also][]{magdis_2012,magnelli_2014,bethermin_2015,faisst_2017}.
        }
\vspace{0.2cm}
\end{figure*}

\subsection{The observed (sub)mm-to-2\,mm colors} 
\label{subsec:beta}
The \textit{observed} 850$\,\mu$m-to-2\,mm and 1.1\,mm-to-2\,mm flux density ratios provide constraints on the nature and dust properties of galaxies.
On the one hand, for galaxies at $z\lesssim2$, these broadbands probe their Rayleigh-Jeans dust emission, providing a measure of their dust emissivity spectral index, $\beta$.
While $\beta$ varies on Galactic scales, extragalactic measurements converge to 1.5$\,\leq\,$$\beta$$\,\leq\,$2.0 \citep[e.g.,][]{Dunne_2001,magnelli_2012}, corresponding to 850$\,\mu$m-to-2\,mm and 1.1\,mm-to-2\,mm flux density ratios in the range 8--11 and 20--30, respectively.
On the other hand, for galaxies at $z\gtrsim2$, the 850$\,\mu$m and 1.1\,mm broadbands probe closer to the peak of the dust emission, yielding much lower 850$\,\mu$m-to-2\,mm and 1.1\,mm-to-2\,mm flux density ratios, which in absence of robust redshift measurements can be used to support their high-redshift nature.

The observed 850$\,\mu$m-to-2\,mm and 1.1\,mm-to-2\,mm flux density ratios of our nine GISMO galaxies as a function of their 1.1\,mm flux densities and, when available, redshifts (see Tab.~\ref{tab:flux infrared}) are displayed in Fig.~\ref{fig:flux ratio}.
For sources with no counterparts in the AzTEC 1.1mm or {\sc Scuba}-2 850$\,\mu$m catalogs, we used as upper limits the 80\% completeness limits of these respective surveys (see Sect.~\ref{sec:counterp}).

The four GISMO galaxies with millimeter counterparts (i.e., GISMO-C1/-C2/-C3/-C7) have a flux density ratio distribution consistent with that observed in the GDF (S14) and in the gravitationally-lensed galaxy sample from the SPT \citep[i.e., South Pole Telescope;][]{strandet_2016}.
This agreement demonstrates the quality of the GISMO calibration but also suggests that our galaxies have intrinsic properties similar to galaxies in those samples, i.e, high-redshift highly star-forming galaxies.
In addition, our galaxies as well as those from the GDF and SPT, have at their redshifts higher 850$\,\mu$m-to-2\,mm and 1.1\,mm-to-2\,mm flux density ratios than predicted from local SED templates.
This implies that high-redshift star-forming galaxies have in average hotter dust temperatures than in the local Universe \citep[see also, e.g.,][]{magdis_2012,magnelli_2014,bethermin_2015,faisst_2017}.

GISMO-C6 has a 1.1\,mm-to-2\,mm upper limit consistent with the rest of the distribution but its very low 850$\,\mu$m-to-2\,mm flux density ratio is at odds with the rest of the distribution if this galaxy is at $z=1.003$.
This suggests that the redshift association of this source is likely incorrect, as already discussed in Sect.~\ref{sec:counterp}. 

All GISMO galaxies without (sub)mm counterparts have low 850$\,\mu$m-to-2\,mm upper limits, consistent with a high-redshift nature. 
Among those, GISMO-C4 exhibits the lowest 850$\,\mu$m-to-2\,mm upper limits.
Combined with its low false detection probability ($\sim\,$6\%), this makes GISMO-C4 a robust high-redshift ($z>4$) candidate.

Finally, as a sanity check, we evaluated lower limits on the 1.1\,mm-to-2\,mm flux density ratios of all AzTEC-1.1\,mm sources within our deep coverage ($\sigma$$\,\sim\,$0.23\,mJy/beam) but undetected by GISMO at S/N$\,\geq\,$3.7.
Here, we used as 2\,mm upper limits, $5\,\times$ the value of the GISMO noise at the position of the AzTEC-1.1\,mm sources.
None of these lower limits are at odds with the rest of the distribution.
This implies that our GISMO survey did not ``miss'' any plausible 2\,mm emitters. 

\subsection{Redshift Distribution}
\label{subsec:redshift}
\begin{figure}
\includegraphics[width=\linewidth]{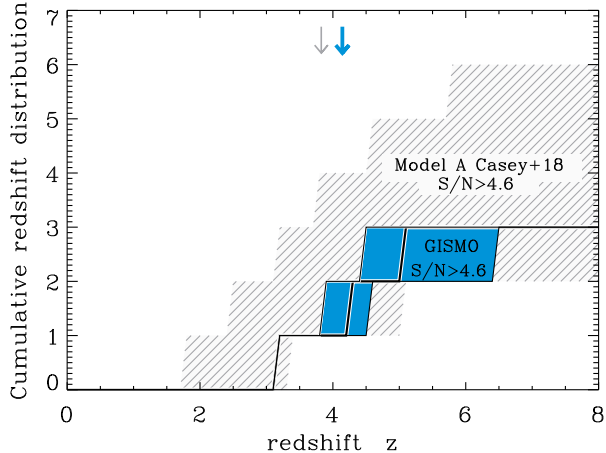}		
\caption{\label{fig:redshift}
		Cumulative redshift distribution of our S/N$\,\geq\,4.6$ (solid line; blue region) GISMO detected galaxies. 
        At this detection significance, all three galaxies have a redshift measurement and thus constitute a \textit{redshift-complete} 2\,mm-bright galaxy sample.
        The lower and upper envelopes of the blue region show the 16th and 86th percentiles of their cumulative redshift distributions using 1,000 Monte-Carlo realizations, drawing the redshift of each sources randomly and uniformly within their redshift uncertainties (Tab.~\ref{tab:flux infrared}).
        The thick blue downward pointing arrow shows the median redshift measured over these 1,000 realizations. 
        The gray hatched regions shows the corresponding cumulative redshift distributions from the model A of \citet{casey_2018}.
        The lower and upper envelopes of these regions represent the 16th and 86th percentiles of these distributions in 1,000 S/N$\,\geq\,4.6$ mock catalogs, i.e., injecting galaxies within our jackknifed map and retrieving them using the same source extraction method as that used to produce our real catalog.
        The thin gray downward pointing arrow shows the median redshift over these 1,000 mock catalogs.}
\end{figure}

Recent observations as well as phenomenological galaxy evolution models suggested that flux density and wavelength selection are crucial determining factors in the redshift distribution of (sub)mm-selected galaxy samples \citep[e.g,][]{younger_2007,younger_2009,bethermin_2015,strandet_2016,brisbin_2017,casey_2018}.
In particular, these studies found that selecting bright galaxies at long wavelengths ($\lambda_{\rm obs}>1.1$mm) provides the most favorable criterion for picking out high-redshift star-forming galaxies.
With our GISMO-2\,mm wide survey, we can further explore these findings.

The three brightest galaxies (i.e., S/N$\,\geq\,$4.6) in our catalog have all a (tentative) redshift measurement:
GISMO-C1 has a spectroscopic redshift of 3.179; GISMO-C2 a (sub)mm-to-radio-based photometric redshift of 3.89$^{+3.11}_{-0.67}$; and GISMO-C3 an optical/NIR-based photometric redshift of $4.58^{+0.25}_{-0.68}$ \citep[][see Sect.~\ref{sec:counterp} and Tab.~\ref{tab:flux infrared}]{brisbin_2017}.
In Fig.~\ref{fig:redshift}, we show the cumulative redshift distribution of this \textit{redshift-complete} 2\,mm-bright (i.e., S/N$\,\geq\,$4.6) galaxy sample.
To account for the large redshift uncertainties for GISMO-C2 and GISMO-C3, we used 1,000 realizations of our catalog, each time drawing the redshift of our sources randomly and uniformly within their redshift uncertainties as given in Tab.~\ref{tab:flux infrared} (see also Sect.~\ref{sec:counterp}).
In Fig.~\ref{fig:redshift}, we plotted the 16th, median and 86th percentiles of these 1,000 cumulative redshift distributions.
The median redshift of our 2\,mm-bright galaxy sample is $\tilde{z}=4.1$, significantly higher than that of the COSMOS AzTEC/ASTE 1.1\,mm sample analyzed by \citet{brisbin_2017}, i.e., $\tilde{z}=2.45$.
However, restricting their sample to the five galaxies with $S_{\rm 1.3\,mm}\geq4.07\,$mJy, (the faintest 1.3\,mm flux density found within our 2\,mm-bright galaxy sample; Tab.~\ref{tab:flux infrared}), their median redshift increases to $\tilde{z}=4.3$, following their conclusion that brighter millimeter sources are preferentially found at higher redshifts.
While based on very small numbers, this suggests that bright sources at 1.1\,mm and 2\,mm surveys yield very similar redshift distribution.
However, the possible advantage of 2\,mm surveys in picking out higher redshift star-forming galaxies than that at 1.1\,mm might only be revealed by probing even larger co-moving volumes. 
In addition, one has to bear in mind that GISMO-C4 -- our next brightest detection -- has no AzTEC-1.1\,mm counterpart and could thus potentially lie at very high redshifts.

In Fig.~\ref{fig:redshift}, we also compare our findings to predictions from the model A of \citet{casey_2018} that best describes our number count measurements (see Sect.~\ref{subsec:counts}). 
To this end, we used mock maps generated from this model, similarly as in Sect.~\ref{subsec:extraction}.
In Fig.~\ref{fig:redshift}, we plotted the 16th and 86th percentiles of the cumulative redshift distributions of sources retrieved with S/N$\,\geq\,$4.6 in 1,000 mock maps.
The redshift distribution as well as the median redshift predicted by this model (i.e., $\tilde{z}_{\rm model}=3.8$) is consistent with our observations.

Unfortunately we could not explore the effect of different flux density cuts on the median redshift of our 2 mm-selected samples.
Indeed, cutting our catalog at lower S/N would include galaxies for which no redshift informations are yet available (i.e., GISMO-C4, GISMO-C5, GISMO-C8 and GISMO-C9).
Such analysis is postponed until further follow-up observations on these galaxies are made.

\subsection{FIR-to-mm Spectral Energy Distribution}
\label{subsec:SED}
\begin{figure*}
	\begin{center}
		\includegraphics[width=\columnwidth]{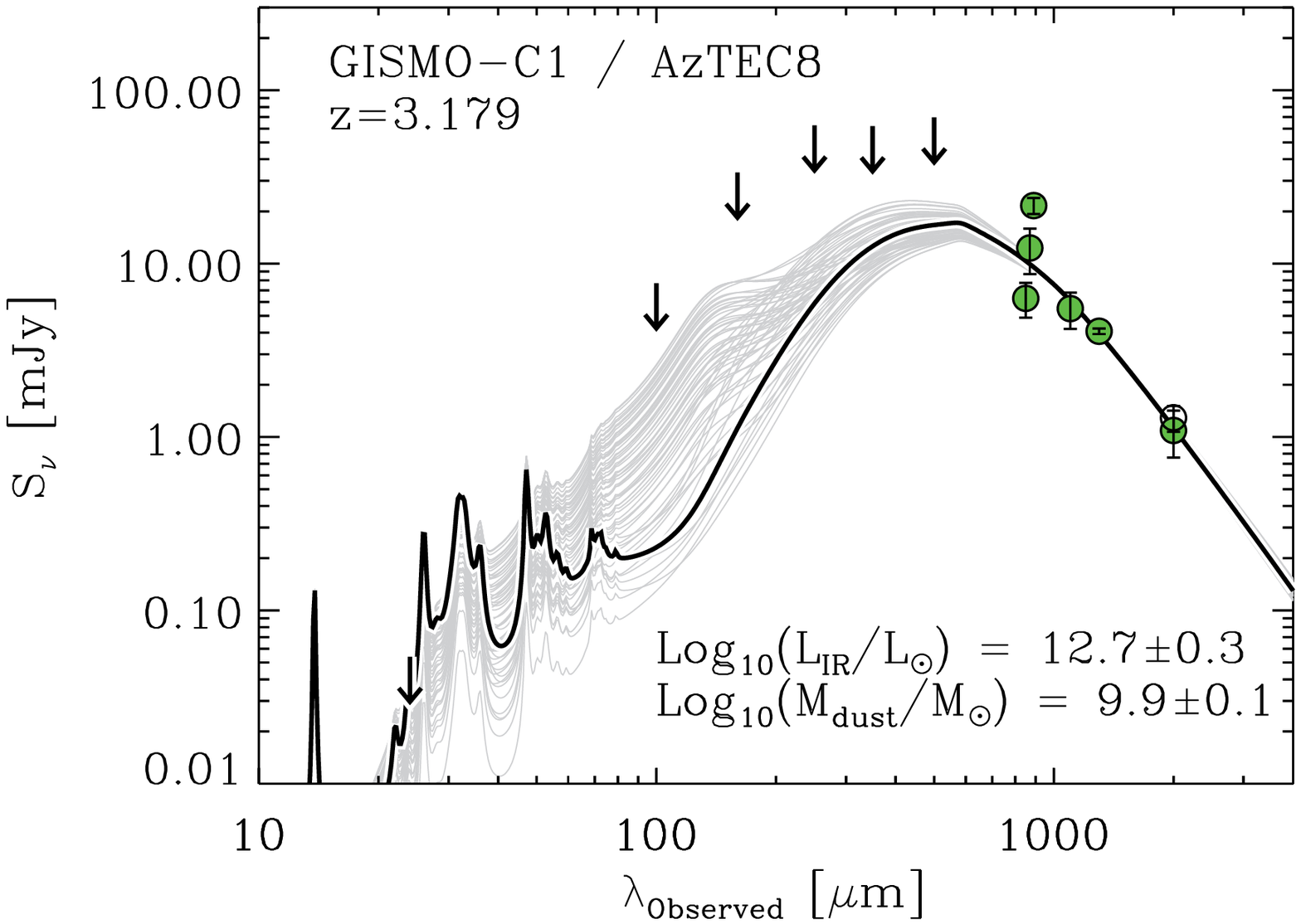}\vspace{0.2cm} 
		\includegraphics[width=\columnwidth]{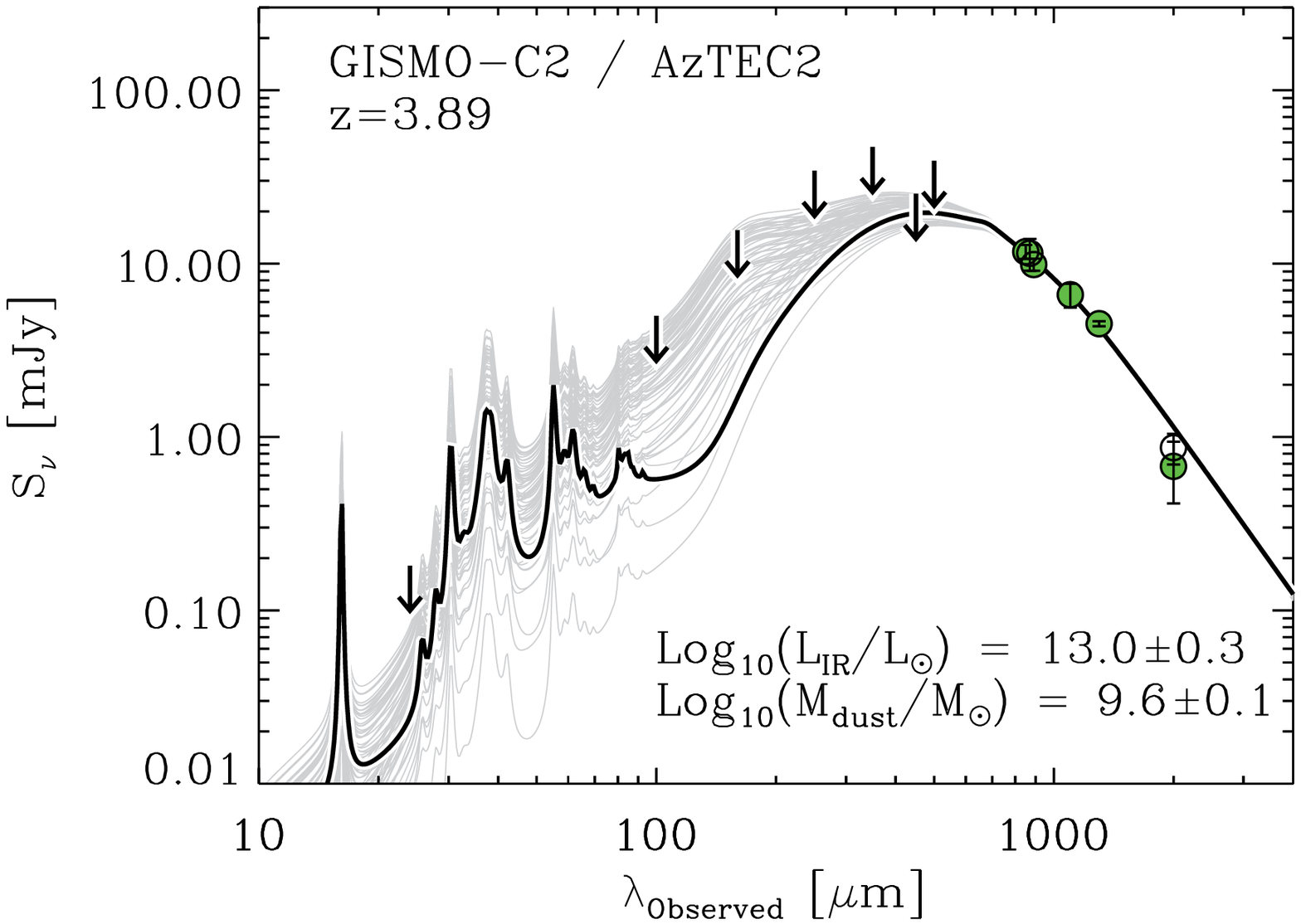}\vspace{0.2cm}
		\includegraphics[width=\columnwidth]{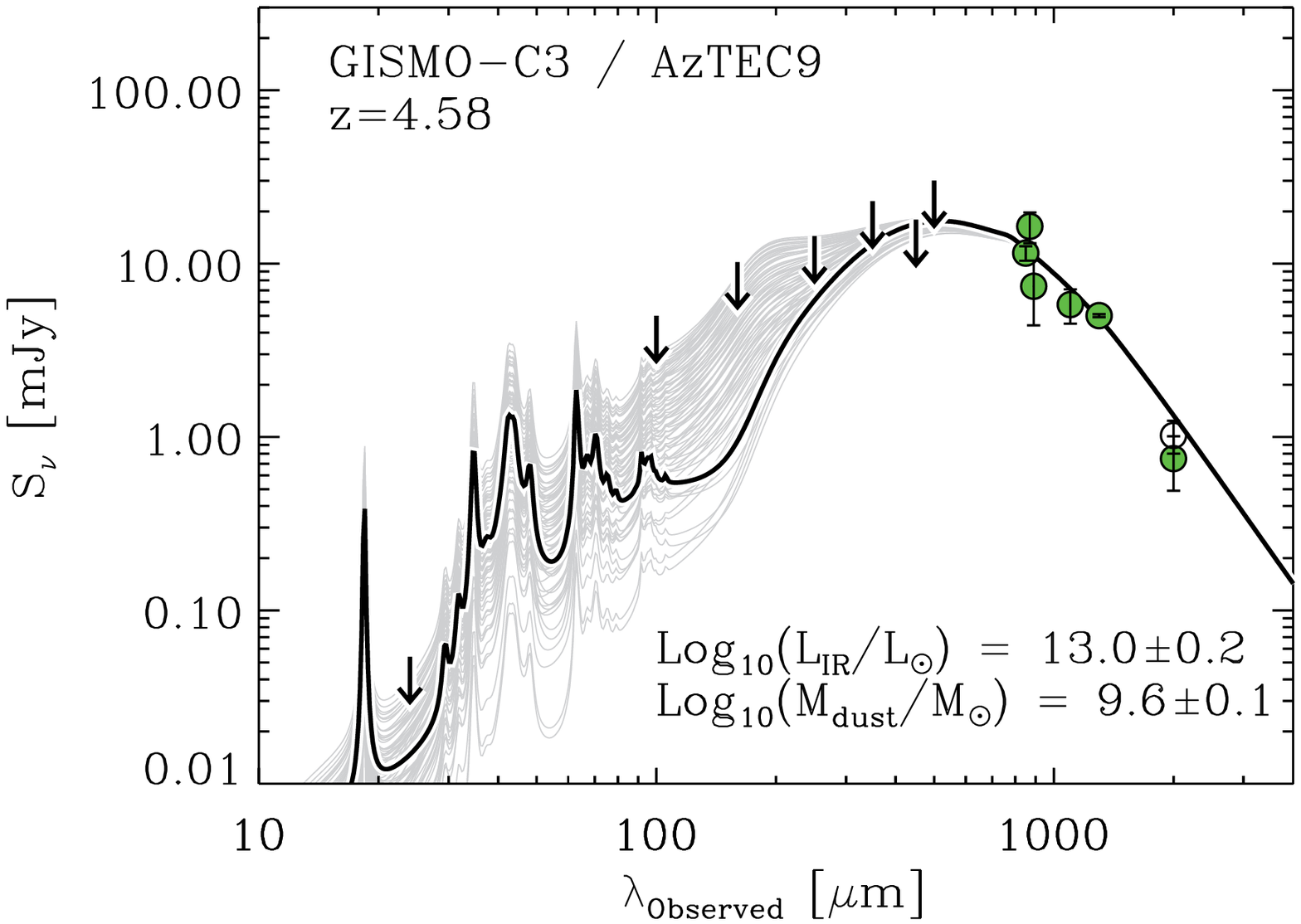}\vspace{0.2cm} 
		\includegraphics[width=\columnwidth]{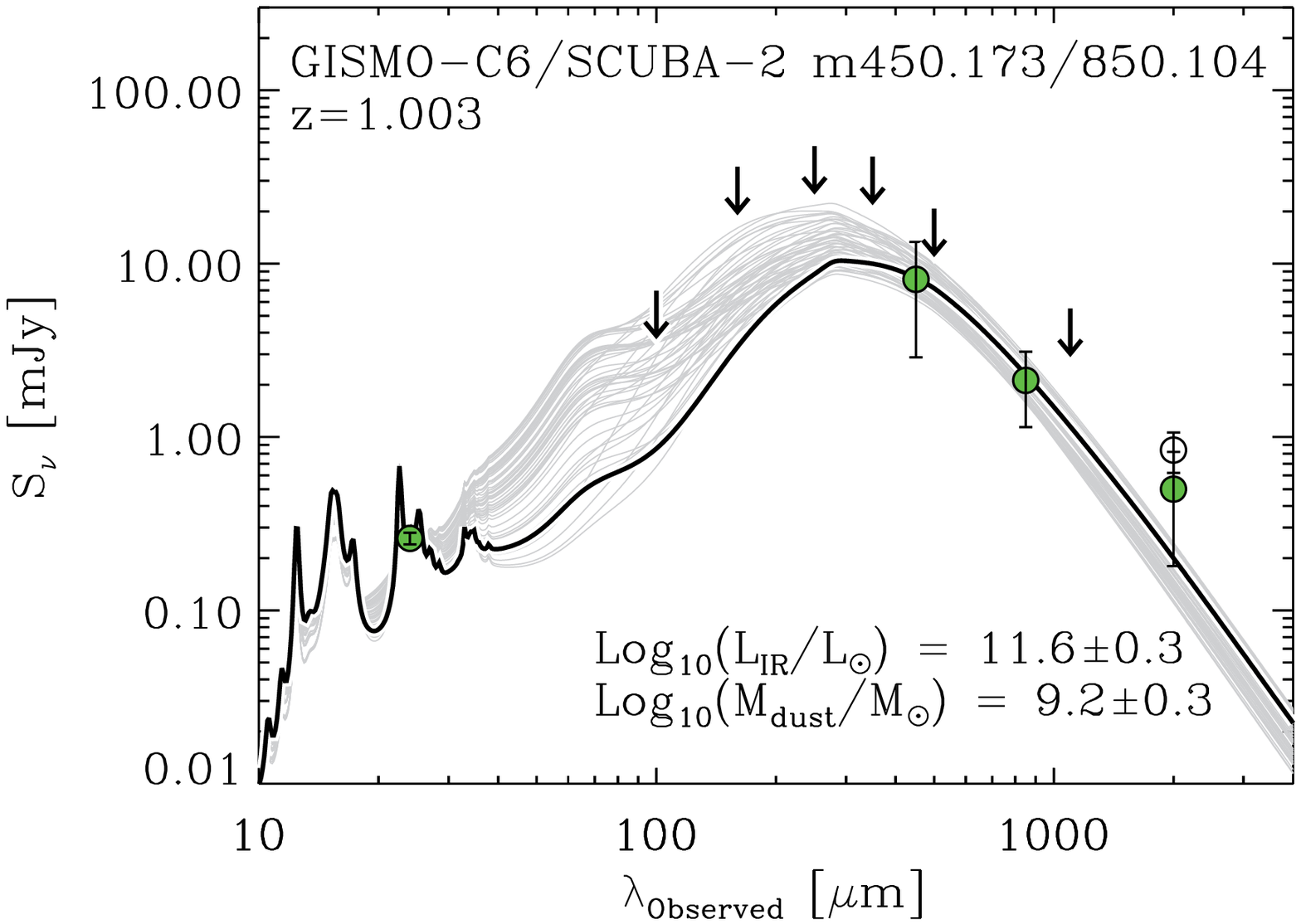}\vspace{0.2cm}
		\includegraphics[width=\columnwidth]{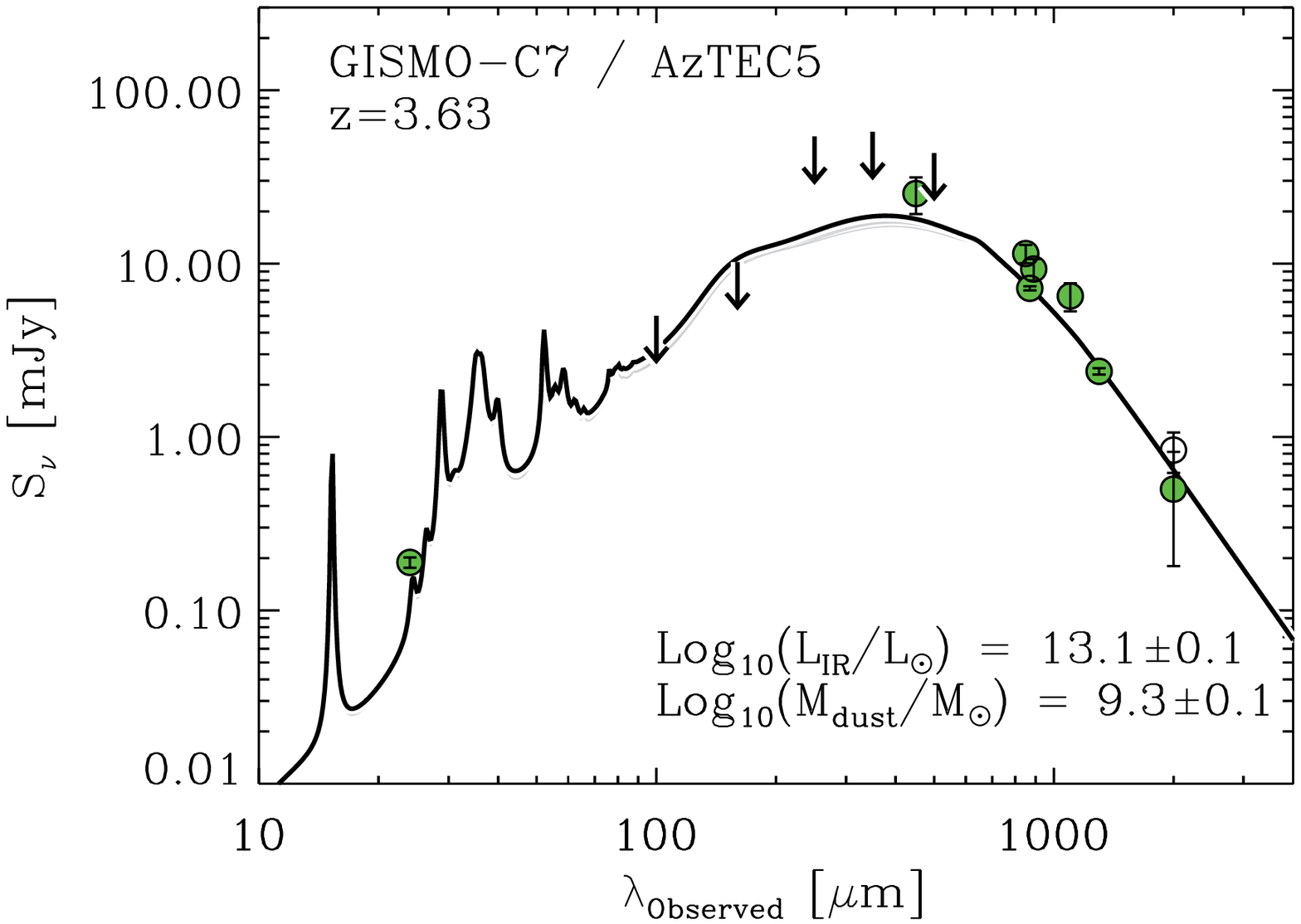}\vspace{0.2cm} 
		\caption{\label{fig:sed}
        Broadband SED of our S/N$\,\geq\,3.7$ GISMO COSMOS sources with known (sub)mm counterparts.
        The best fit \citet{draine_2007} model is shown by the black thick line. 
        Light gray lines present the range of \citet{draine_2007} models with $\chi^{2}_{\rm redu}< {\rm min}(\chi^{2}_{\rm redu})+1$.
        Flux densities (green filled circles) and upper limits (downward pointing arrows) used in these fits are given in Tab.~\ref{tab:flux infrared}.
        At 2\,mm, we also show as opened circles the \textit{observed} GISMO flux densities, however, not used in these fits.
        }
	\end{center}
\end{figure*}
We derived the infrared luminosities, SFRs and dust masses ($M_{\rm dust}$) of all GISMO galaxies with tentative redshift measurements (i.e., GISMO-C1, GISMO-C2, GISMO-C3, GISMO-C6 and GISMO-C7) via mid-infrared-to-(sub)mm SED fitting using the \citet{draine_2007} dust model\footnote{The low number of models compatible with GISMO-C7 is due to its particular photometry. It has a high-significance MIPS-24$\,\mu$m detection and very constraining PACS-100$\,\mu$m and -160$\,\mu$m upper limits with respect to its (sub)mm flux densities.} (Fig.~\ref{fig:sed}; Tab.~\ref{tab:flux infrared}).
The mid-infrared-to-(sub)mm photometry and redshift used in these fits are summarized in Tab.~\ref{tab:flux infrared} and were discussed in Sect.~\ref{sec:counterp}.

As suggested by the observed (sub)mm-to-2\,mm colors (Sect.~\ref{subsec:beta}), the 2\,mm flux densities measured by GISMO are consistent with the overall SEDs of these galaxies while putting additional constraints on their Rayleigh-Jeans dust emission.
Only GISMO-C6 exhibits an unusual dust SED, which peaks at very long wavelength ($\lambda_{\rm rest}^{\rm peak}$$\,\sim\,$150$\,\mu$m), corresponding to a low luminosity-weighted dust temperature of $\sim\,$20\,K.
This low dust temperature could be explained by the well-known $L_{\rm IR}-T_{\rm dust}$ selection bias affecting (sub)mm surveys \citep[e.g.,][]{magnelli_2012}.
However, this could also suggest that the redshift association for this source is incorrect, as already discussed in Sect.~\ref{sec:counterp}. 
\begin{figure}
\includegraphics[width=\linewidth]{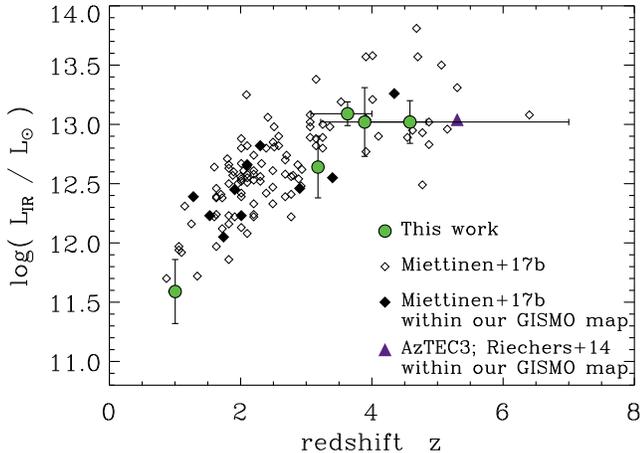}		
\caption{\label{fig:lir_redshift}
		The redshift-infrared luminosity distribution of the five GISMO COSMOS galaxies with (sub)mm counterparts and redshift measurements (green filled circles).
        Diamonds show the redshift-infrared luminosity distribution of all AzTEC/ASTE sources in \citet{miettinen_2017c} but those associated with a GISMO counterpart (i.e., AzTEC/C2, AzTEC/C3, AzTEC/C14 and AzTEC/C42).
     	Filled diamonds correspond to AzTEC/ASTE sources not detected at S/N$\,\geq\,$3.7 in the GISMO map, while opened diamonds correspond to AzTEC/ASTE sources not covered by our survey.
     	The blue triangle corresponds to AzTEC3, a high-redshift starburst not detected at S/N$\,\geq\,$3.7 in the GISMO map. 
     	While this source is bright in the AzTEC/JCMT map, it is relatively faint the AzTEC/ASTE map (a.k.a. AzTEC/C138).
     	It was thus not included in \citet{miettinen_2017c}, which studied AzTEC/ASTE sources up to AzTEC/C129.}
\vspace{0.2cm}
\end{figure}

All our galaxies except GISMO-C6, are very luminous ($L_{\rm IR}$$\,>\,$$10^{12.6}\,$L$_\odot$), corresponding to SFRs in the range 400 -- 1200\,M$_{\odot}\,$yr$^{-1}$; assuming a Chabrier IMF and the relation SFR\,[M$_{\odot}\,$yr$^{-1}$]$\,=\,10^{-10}\times L_{\rm IR}\,$[L$_\odot$] \citep{kennicutt_1998}.
Even with their relatively large stellar masses ($10^{10.8-11.5}\,$M$_{\odot}$; Tab.~\ref{tab:flux infrared}), these high SFRs locate these galaxies on the upper-part or above the $z$$\,\sim\,$$4$ main-sequence of star-forming galaxies \citep[e.g.,][]{schreiber_2015}. 
These are thus ``starburst'' galaxies with respect to the bulk of the star-forming galaxy population at these redshifts.

When compared to the AzTEC-1.1\,mm-selected galaxies within our map \citep[Fig.~\ref{fig:lir_redshift};][]{miettinen_2017c}, it becomes clear that our GISMO 2\,mm survey picked out the brightest and highest redshift galaxies among them.
Among the six AzTEC-1.1\,mm sources within our map with $L_{\rm IR}$$>\,$10$^{12.6}\,$L$_{\odot}$ and $z>3$, four are detected by our GISMO survey (i.e., 66\%).
The only AzTEC-1.1\,mm-selected galaxy not detected by our survey (S/N$\,\geq\,$3.7) and with $L_{\rm IR}>10^{12.6}\,$L$_\odot$ and $z>3$ are AzTEC1 (a.k.a. AzTEC/C5) and AzTEC3 (a.k.a. AzTEC/C138).
However, the GISMO flux density upper limits for these galaxies yield 1.1\,mm-to-2\,mm lower limits consistent with the rest of the distribution (see the black upward pointing arrows with $S_{1.1{\rm mm}}=9.3$\,mJy and 5.9\,mJy for AzTEC1 and AzTEC3, respectively, in the bottom-left panel of Fig.~\ref{fig:flux ratio}).
These non-detections could thus be simply explained by the inherent incompleteness of our catalog at faint flux densities (see Fig.~\ref{fig:Completeness}) or by particularly hot dust emission, especially in the case of AzTEC3 \citep{riechers_2014}.
In the GISMO map, at the position of AzTEC1, we find a S/N$\,\sim\,$3.4 detection with $S_{\rm 2\,mm}=0.78\pm0.23\,$mJy, while at the position of AzTEC3 we find a S/N$\,\sim\,$2.4 detection with $S_{\rm 2\,mm}=0.53\pm0.22\,$mJy.

We measured massive dust content in all GISMO-detected galaxies ($10^{9.3-9.9}\,$M$_{\odot}$; Tab.~\ref{tab:flux infrared}).
From the dust and stellar masses of these galaxies, we can calculate the required dust yields per AGB star and per SNe, following \citet{michalowski_2010b,michalowski_2010}, i.e.,
\begin{equation}
    M_{\rm dust}/N(M_0<M<M_1),
\end{equation}
where $N(M_0<z<M_1)$ is the number of stars with masses between $M_0$ and $M_1$ in the stellar population with a total mass of $M_\ast$, i.e., 
\begin{equation}
    N(M_0<M<M_1) = M_\ast \frac{\int_{M_0}^{M_1} {\rm IMF}(M)\,dM}{\int_{M_{\rm min}}^{M_{\rm max}} {\rm IMF}(M)\,M\,dM},
\end{equation}
and where IMF$(M)$ is the initial mass function from \citet{chabrier_2003} with $M_{\rm min}=0.15\,$M$_\odot$ and $M_{\rm max}=120\,$M$_\odot$.
As in \citet{michalowski_2010b}, for AGB stars we assumed $M_0=2.5\,$M$_\odot$ and $M_1=8\,$M$_\odot$, whereas for SNe we assumed $M_0=8\,$M$_\odot$ and $M_1=120\,$M$_\odot$.
We measure dust yields per AGB star of 1.9$\pm$0.7, 1.3$\pm$0.5 and 0.15$\pm$0.05\,M$_\odot$ for GISMO-C1, GISMO-C3 and GISMO-C7, respectively, while theoretical works predict dust yields $\lesssim\,$$4\times10^{-2}\,$M$_\odot$ \citep[][and reference thererin]{michalowski_2010b,michalowski_2010}.
We calcutate dust yields per SNe of 6.9$\pm$2.5, 4.9$\pm$1.8 and 0.56$\pm$0.21\,M$_\odot$ for GISMO-C1, GISMO-C3 and GISMO-C7, respectively.
As for AGB stars, these yields are in tension with theoretical expectations which are $\lesssim\,$$1.32\,$M$_\odot$ without dust destruction and $\lesssim\,$$0.1\,$M$_\odot$ with dust destruction \citep[][and reference thererin]{michalowski_2010b,michalowski_2010}.
As pointed out in \citet{michalowski_2010b,michalowski_2010}, such unrealistically high dust yields suggest efficient dust production in the interstellar medium of these galaxies.
Note that GISMO-C2 is excluded from this analysis because it has no stellar mass estimate.
Indeed, its optical/NIR photometry remains very uncertain as it appears very obscured and unfortunately situated in the vicinity of a bright optical/NIR foreground galaxy (see Sect.~\ref{sec:counterp} and Fig.~\ref{fig:counterparts}).
GISMO-C6 is also excluded from this analysis because of its most probably wrong redshift identification (see Sect.~\ref{sec:counterp}).

Assuming a standard gas-to-dust ratio of 100 appropriate for massive systems \citep[][]{leroy_2011}, these dust masses also translate into large gas reservoirs  several time more massive than the stellar component of these galaxies.
Yet, these gas-rich galaxies with their extreme star-formation activities deplete these reservoirs in $<$1--2\,Gyr, in agreement with lower redshift observations \citep[e.g.,][]{tacconi_2018}.
GISMO-C6 is again the only galaxy with an usually large depletion time scale of $5.7^{+7.3}_{-3.8}$\,Gyr.

Note that deriving the gas masses of these galaxies using the methodology advocated in \citet{scoville_2016} (i.e., from our 2\,mm deboosted flux densities and their Eq.~16), yields measurements in perfect agreement (within $\sim\,$0.1\,dex) with those inferred from the dust model of \citet{draine_2007} and a gas-to-dust ratio of 100.
The only exception is again GISMO-C6 for which the methodology of \citet{scoville_2016} leads to $\sim\,$0.4\,dex lower gas mass.
\begin{figure*}
\includegraphics[width=0.5\linewidth]{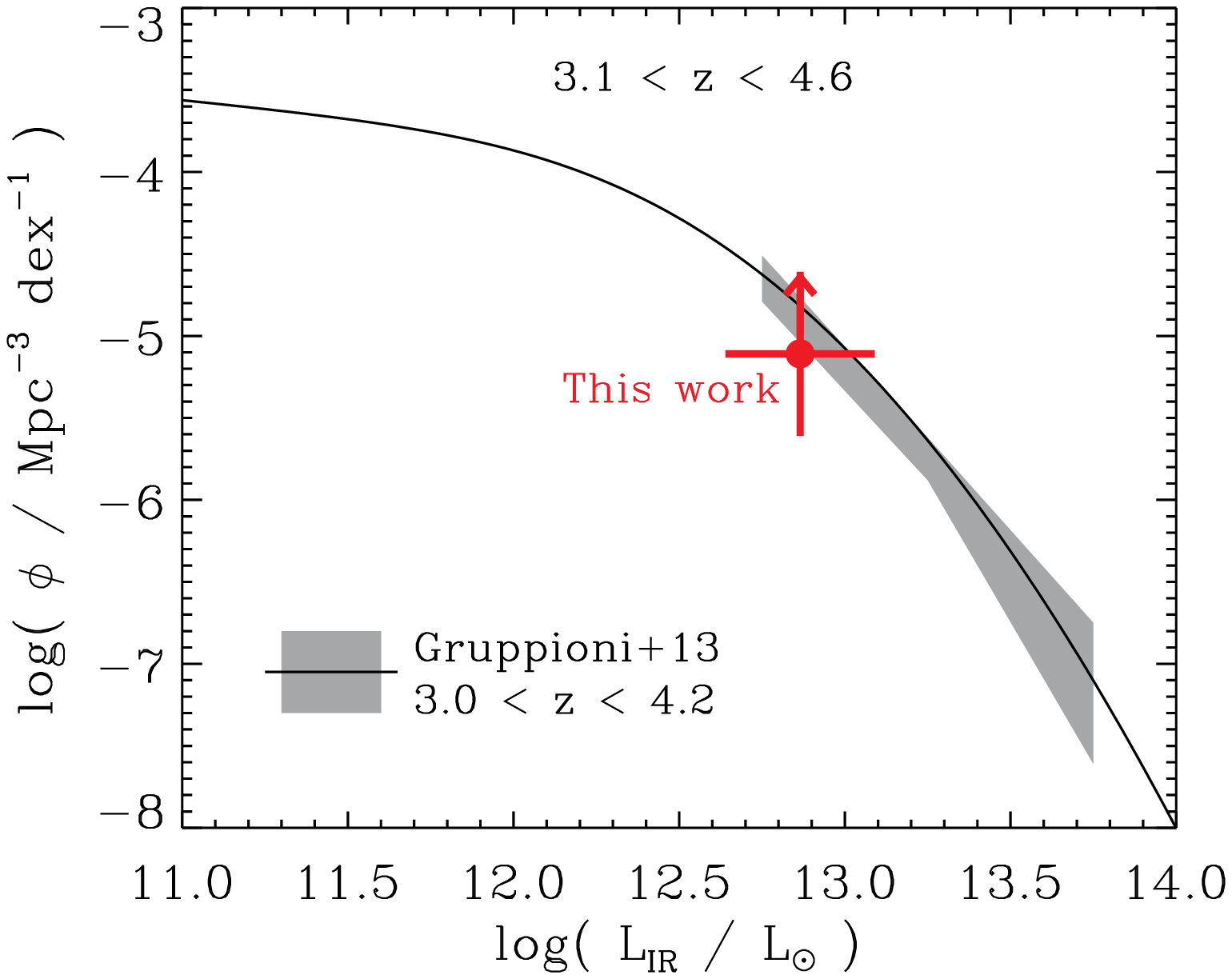}	
\includegraphics[width=0.5\linewidth]{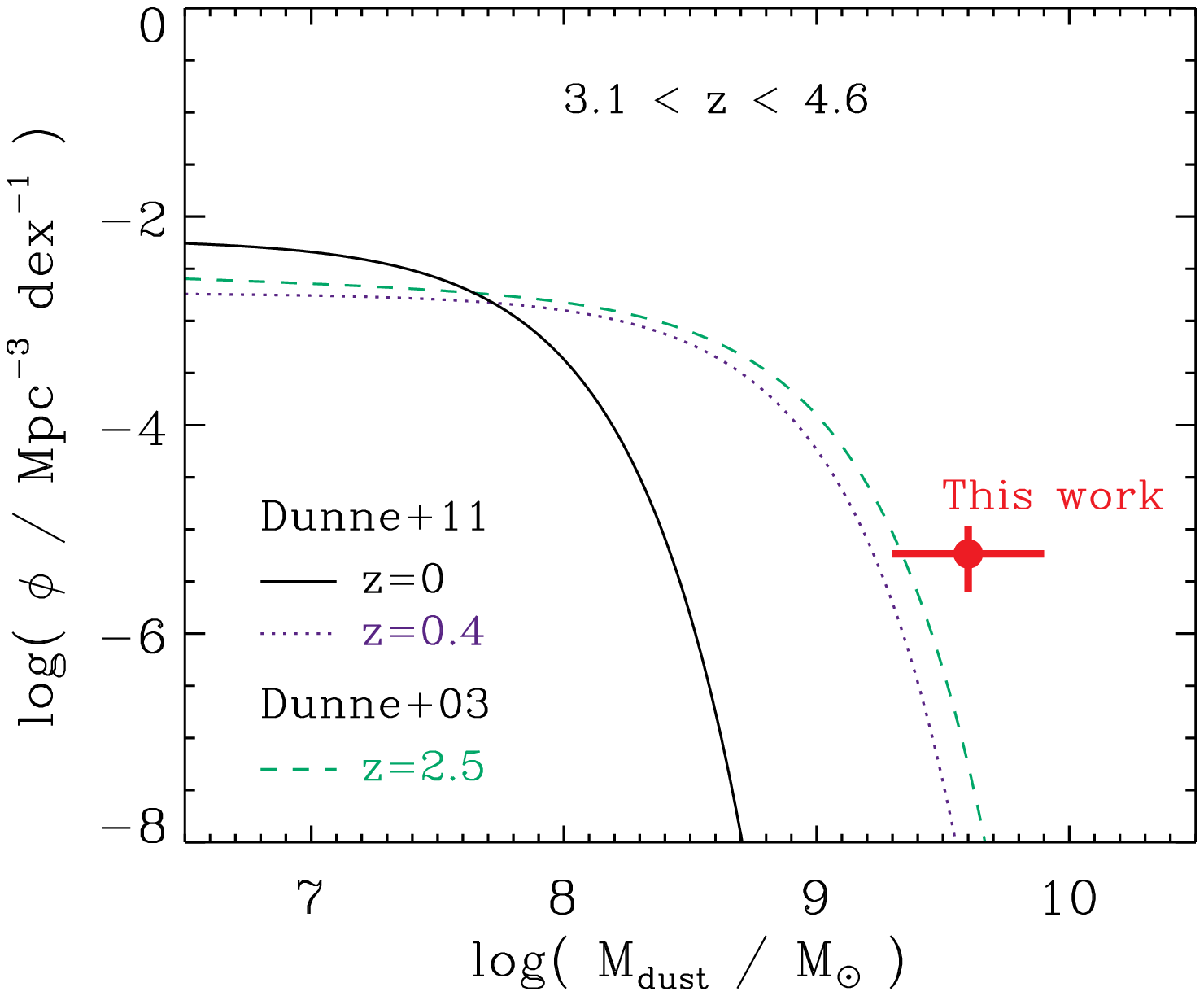}	
\includegraphics[width=0.5\linewidth]{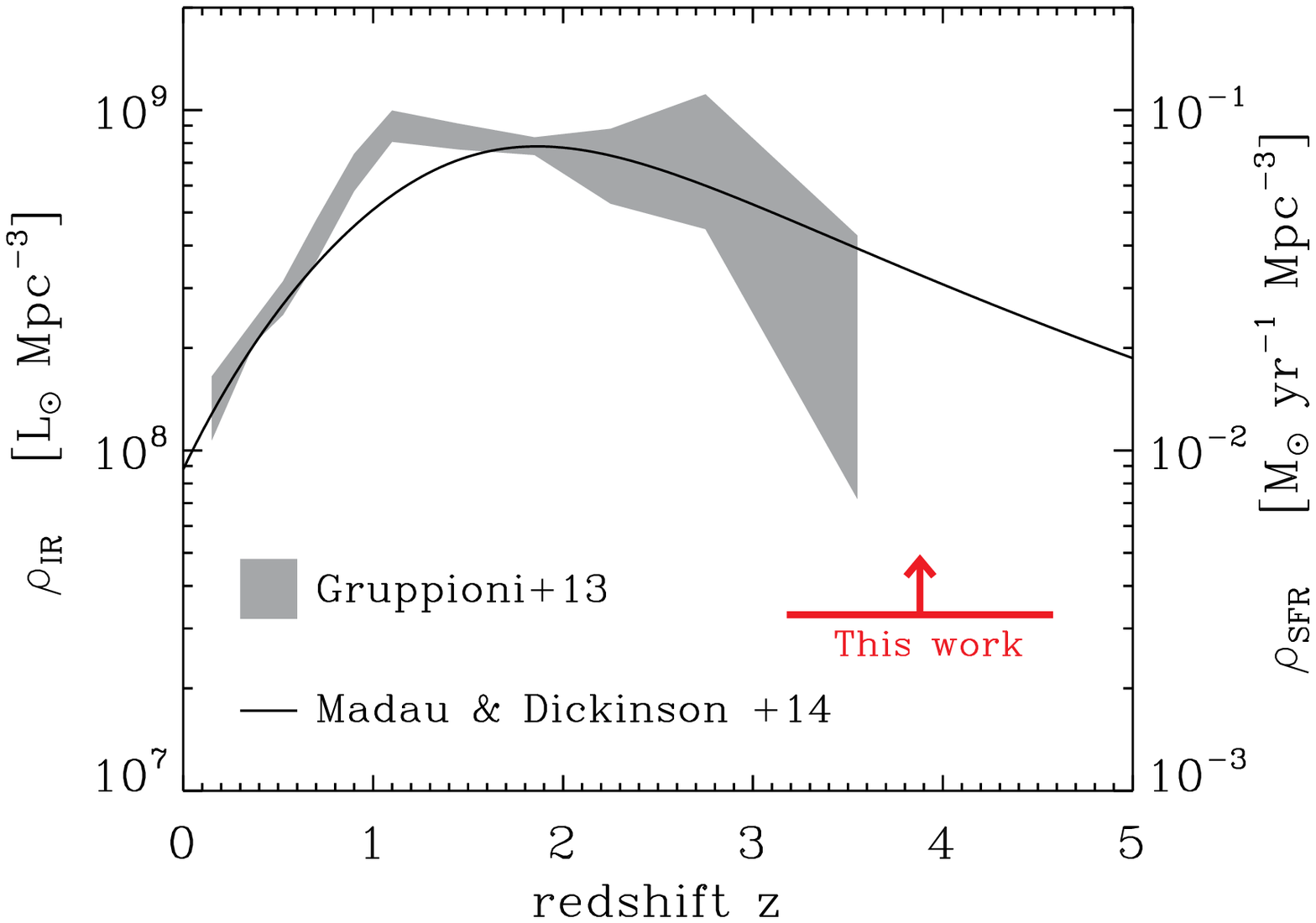}	
\includegraphics[width=0.5\linewidth]{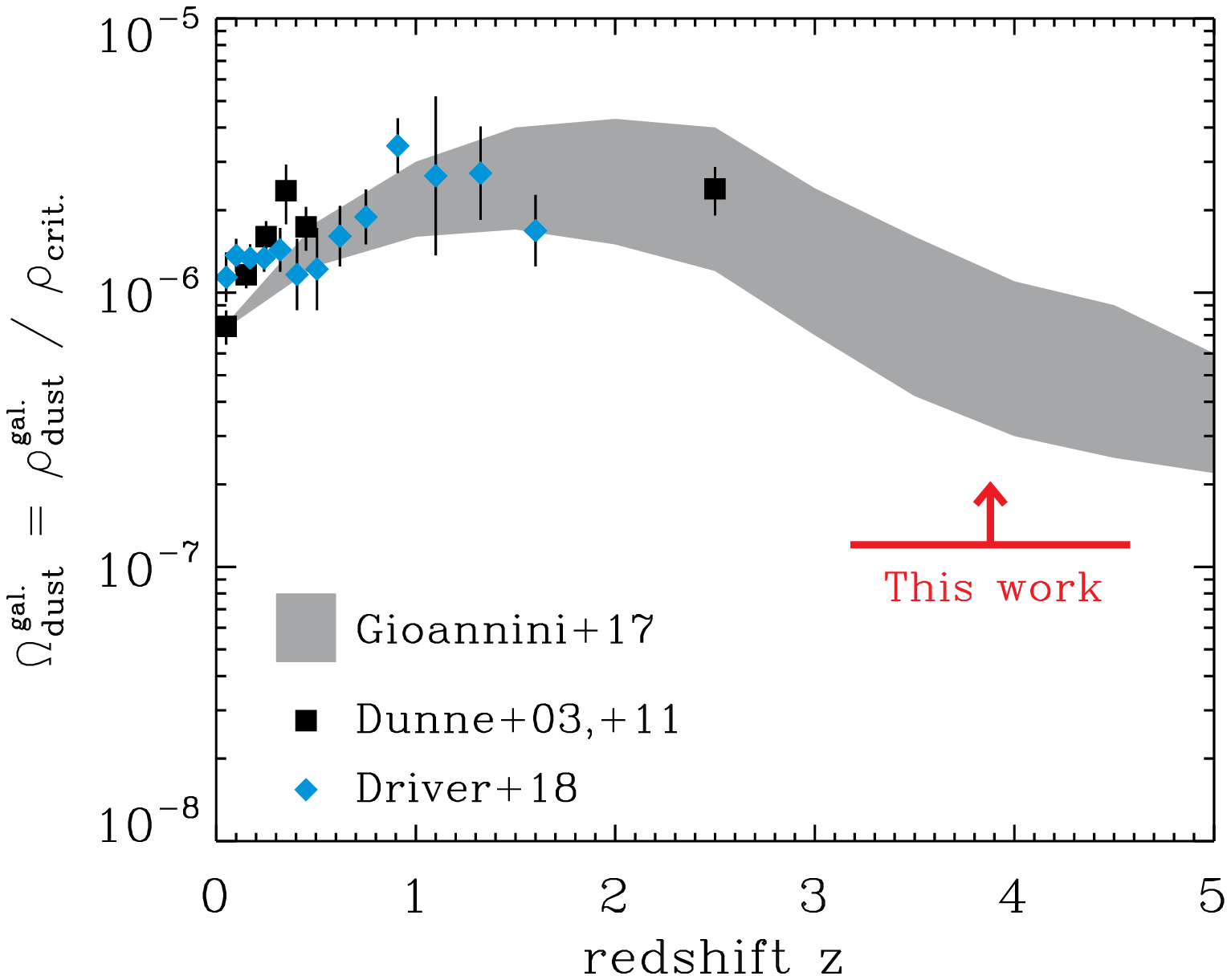}	
\caption{\label{fig:Omega_dust}
        (\textit{Upper left}) Infrared luminosity (i.e., $L_{\rm IR}$[8-1000\,$\mu$m]) function as derived from our four GISMO COSMOS galaxies at $3.1<z<4.6$, i.e., GISMO-C1/-C2/-C3/C7 (red upward pointing arrow).
        The grey region and black solid line show the results from \citet{gruppioni_2013} at $3.1<z<4.6$, obtained from wide COSMOS surveys performed by \textit{Herschel}.
        (\textit{Lower left}) Redshift evolution of the cosmic infrared luminosity density.
        Our high-redshift upper limit is shown by a red upward pointing arrow and corresponds to the sum of the infrared luminosities of GISMO-C1/-C2/-C3/C7 divided by the comoving volume within $3.1<z<4.6$.
        The grey region shows the measurements from \citet{gruppioni_2013}.
        The cosmic infrared luminosity density can be translated into cosmic SFR density (see the right-end y-axis), assuming a Chabrier IMF and the relation SFR\,[M$_{\odot}\,$yr$^{-1}$]$\,=\,10^{-10}\times L_{\rm IR}\,$[L$_\odot$] \citep{kennicutt_1998}.
        The redshift evolution of the cosmic SFR density constrained with a plethora of obscured and unobscured SFR indicators and parametrized in \citet{madau_2014}, is shown by the black solid line.
        (\textit{Upper right}) Dust mass function in galaxies as derived from our four GISMO COSMOS galaxies at $3.1<z<4.6$ (red circle).
        The redshift evolution of the dust mass function in galaxies measured by \citet{dunne_2011} at $z=0$ and $z=0.4$, and \citet{dunne_2003} at $z=2.5$ are shown by the black solid, blue dotted and green dashed lines, respectively.
        (\textit{Lower right}) Redshift evolution of the cosmic dust mass density in galaxies, i.e., $\Omega_{\rm dust}^{\rm gal.}=\rho_{\rm dust}^{\rm gal.}/\rho_{\rm crit.}$, where $\rho_{\rm crit.}$ is the critical density of the Universe with $\rho_{\rm crit.}=1.3\times10^{11}\,$M$_{\odot}\,$Mpc$^{-3}$ \citep{planck_cosmo_2016}.
        Our high-redshift upper limit is shown by a red upward pointing arrow and corresponds to the sum of the dust mass reservoir of GISMO-C1/-C2/-C3/C7 divided by $\rho_{\rm crit.}$ and the comoving volume within $3.1<z<4.6$.
        Black squares show results from \citet{dunne_2003,dunne_2011} obtained using far-infrared/(sub)mm observations, while blue diamonds correspond to those from \citet{driver_2018} using a optical-to-far-infrared energy balance approach in the GAMA fields.
        Theoretical predictions from the chemical evolution models of galaxies of \citet{gioannini_2017} are shown by the grey region.
		}
\vspace{0.2cm}
\end{figure*}

\subsection{Cosmic infrared luminosity and dust mass densities of galaxies}
\label{subsec:cosmic density}
Using our high-redshift sample, we constrained the bright-end of the infrared luminosity function and cosmic infrared luminosity density at $z$$\,\sim\,$$4$; and in a pioneer effort, the massive-end of the dust mass function and cosmic dust mass density in galaxies at $z$$\,\sim\,$$4$ (Fig.~\ref{fig:Omega_dust}).
To this end, we summed up the contributions of GISMO-C1, -C2, -C3 and -C7, and considered as comoving volume that probed by our survey (250\,arcmin$^2$) between $z=3.1$ and $z=4.6$.

Our constraint on the $z$$\,\sim\,$$4$ infrared luminosity function (i.e., $\phi$$\,>\,$$2.5$$\,\times\,$$10^{-6}\ $Mpc$^{-3}$\,dex$^{-1}$ at $L_{\rm IR}=10^{12.85}\,$L$_{\odot}$) is consistent with that obtained by \citet[][upper-left panel of Fig.~\ref{fig:Omega_dust}]{gruppioni_2013} using far-infrared (100--500$\,\mu$m) observations from the \textit{Herschel Space Observatory}.
This implies that even though our Rayleigh-Jeans 2\,mm-selection does not in principle provide a luminosity-limited sample, the population of luminous galaxies with hot dust emission missed by this selection (e.g., AzTEC3) does not dominate at $z$$\,\sim\,$$4$.

Our study adds also an interesting lower limit to the cosmic infrared luminosity density at $z$$\,\sim\,$$4$ (lower-left panel of Fig.~\ref{fig:Omega_dust}; $\rho_{\rm IR}$$\,>\,$$3.3$$\,\times\,$$10^{7}$\ L$_{\odot}$\,Mpc$^{-3}$ at $z=3.9$), which to date remains uncertain owing to the sensitivity limits of current far-infrared surveys \citep{madau_2014}.
Assuming that about half of the cosmic star formation rate density inferred at these redshifts from UV-selected surveys should be seen in form of infrared emission \citep[e.g.][]{cucciati_2012,madau_2014}, we estimate that our sample contribute for $\sim\,$20\% of the cosmic infrared luminosity density expected at $z$$\,\sim\,$$4$.

Our pioneer measurement of the $z$$\,\sim\,$$4$ dust mass function of galaxies (upper-right panel of Fig.~\ref{fig:Omega_dust}; $\phi$$\,>\,$$5.8^{+4.9}_{-3.3}$$\,\times\,$$10^{-6}\ $Mpc$^{-3}$\,dex$^{-1}$ at $M_{\rm dust}=10^{9.6}\,$M$_{\odot}$) is consistent with that inferred at $z$$\,\sim\,$$2.5$ in \citet{dunne_2003} by estimating the dust masses of submillimeter galaxies assuming a flat redshift distribution ranging from  $z$$\,=\,$$1$ to $z$$\,=\,$$5$.
Both studies suggest a mild evolution of the massive-end of the dust mass function of galaxies from $z$$\,\sim\,$$0.4$ to $z$$\,\sim\,$$4$.

Finally, we provide a lower limit on the cosmic dust mass density in galaxies at $z$$\,\sim\,$$4$ (lower-right panel of Fig.~\ref{fig:Omega_dust}; $\rho_{\rm dust}^{\rm gal.}$$\,>\,$$1.6$$\,\times\,$$10^{4}$\ M$_{\odot}$\,Mpc$^{-3}$ at $z=3.9$).
This estimate is consistent with recent theoretical expectations, even though theory predicts a drastic decrease of the cosmic dust mass density in galaxies from $z$$\,=\,$$2$ to $z$$\,=\,$$5$ \citep{gioannini_2017}.

\section{Conclusions}
\label{sec:conclusion}
With the GISMO array at the IRAM 30 m-telescope, we performed the widest deep 2\,mm survey to date, reaching a uniform $\sigma$$\,\sim\,$0.23\,mJy\,beam$^{-1}$ sensitivity over an area of $\sim\,$$250$\,arcmin$^2$ in the COSMOS field.
Within this map, we detected four sources with a high detection significance (i.e., S/N$\,\geq\,$4.4), corresponding to a low false detection rate per map of only 0.09 sources.
Five sources detected with 4.4$\,>\,$S/N$\,\geq\,$3.7 are also added to our catalog, among which 1.65 are supposed to be false detections.
With this catalog in hand, we found that:
\begin{itemize}
\item Combined with the GISMO deep field ($\sigma$$\,\sim\,$$0.135$\,mJy\,beam$^{-1}$ over 13\,arcmin$^2$; S14), it provides robust and consistent measurements of the 2\,mm number counts over one decade in flux densities. These give critical constraints for current and upcoming galaxy evolution models. For example, while model A of \citet{casey_2018}, which represents a ``dust-poor'' early Universe, best describes our 2\,mm number counts, we can begin to rule out subsets of their model B ``dust-rich'' Universe, whereby only cutoff redshifts $z_{\rm cutoff}<7$ are consistent with our data.
\item Five sources in our map have counterparts in other deep (sub)mm catalogs available for the COSMOS field \citep{scott_2008,aretxaga_2011,casey_2013,geach_2017}. The redshifts of these sources found in the literature suggest that all but one lie above $z$$\,\sim\,$3. For these four high-redshift galaxies, their GISMO-2\,mm flux densities are consistent with their overall mid-to-far-infrared SEDs, while providing additional constraints on their Rayleigh-Jeans dust emission. These high-redshift galaxies are found to be ultra-luminous infrared galaxies with SFRs in the range 400 -- 1200\,M$_{\odot}\,$yr$^{-1}$. They are associated with large dust/gas reservoirs but their extreme SFRs yield gas depletion time scales of $<$1--2\,Gyr, in agreement with lower redshift observations \citep[e.g.,][]{tacconi_2018}. GISMO-C6 is the only galaxy at odds with this picture, as it is a relatively low redshift galaxy ($z=1$) with a moderate infrared luminosity ($L_{\rm IR}$$\,=\,10^{11.6}\,$L$_\odot$) and a very cold luminosity-weighted dust temperature ($\sim\,$20\,K). The detection of this galaxy by GISMO could be explained by the well-known $L_{\rm IR}-T_{\rm dust}$ selection bias affecting (sub)mm surveys. However, this could also suggest that the redshift association for this source is incorrect.
\item Comparing the redshift-infrared luminosity distribution of our galaxies to that of the AzTEC-1.1\,mm-selected galaxies within our map \citep{miettinen_2017c,riechers_2014}, we found that our GISMO-2\,mm survey is picking out a relatively complete sample ($\sim\,66\%$) of the most luminous ($L_{\rm IR}$$>\,$10$^{12.6}\,$L$_{\odot}$) and highest redshift ($z>3$) galaxies among them. This suggests that the selection of bright galaxies at long wavelengths provides the most favorable criterion for finding massive vigorously star-forming high-redshift galaxies, in agreement with recent observations and galaxy evolution models \citep[e.g.,][]{younger_2007,younger_2009,strandet_2016,brisbin_2017,casey_2018,zavala_2018}. Unfortunately, due to small number statistics, it is not yet possible to fully  quantify the observational advantage of 2\,mm over 1.1\,mm selection for high-redshift studies.
\item GISMO-C4 is the fourth brightest source in our catalog and is thus very unlikely to be a false detection ($P_{\rm f}$$\,=\,$$6.2\%$). Yet, it has no (sub)mm counterpart. Such a (sub)mm dropout could be an unidentified very high-redshift galaxies ($z>4$), as suggested by its unusually low 850$\,\mu$m-to-2\,mm flux density ratio. This very high-redshift candidate will require future dedicated follow-up with ALMA or NOEMA. Three other sources in our catalog have potentially no (sub)mm counterparts and low 850$\,\mu$m-to-2\,mm flux density ratios, however, 1.22 sources are supposed to be false detections. 
\end{itemize}
Our wide GISMO 2\,mm survey, combined with the pencil-beam confusion-limited GISMO deep field (S14), has unambiguously demonstrated the advantage of long wavelength surveys for studying the rare massive high-redshift highly star-forming galaxies.
Such surveys provide valuable constraints on the yet very uncertain bright-end of the infrared luminosity function and massive-end of the dust mass function at $z$$\,\sim\,$$4$.
However, our 2\,mm surveys are still limited by their relatively small sky coverage. The ALMA 2\,mm continuum survey in COSMOS (PI Casey, Cycle 6) combined with current and future 2\,mm instruments on single dish facilities -- like NIKA-II on the IRAM 30m and GISMO-2 and TolTEC on the LMT 50m -- will certainly demonstrate further the utility of long-wavelength selection in solving for the relative abundance of dusty star-forming galaxies in the $z>3$ Universe by mapping large areas of sky to sub-mJy depths.
They will provide thereby invaluable samples of efficiently-selected high-redshift sources for interferometric spectral scan follow-up with ALMA and NOEMA, enabling the study of dust production within the first $\sim$2\,Gyr of cosmic time.

\begin{table*}
\begin{center} 
\caption{\label{tab:flux infrared} GISMO (sub)mm counterparts}
\begin{tabular}{ c c c c c c | c} 
\hline \hline
\rule{0pt}{3ex}			         & GISMO-C1           & GISMO-C2          & GISMO-C3          & GISMO-C6             & GISMO-C7  &  GISMO-C4$^{\rm{o}}$\\ [0.5ex]
\hline
\rule{0pt}{5ex}        & AzTEC8   & AzTEC2  & AzTEC9 & m450.173/850.104 & AzTEC5 \\
\rule{0pt}{3ex}(Sub-)mm name       & AzTEC/C2   & AzTEC/C3  & AzTEC/C14 &   & AzTEC/C42 & \dots \\
\rule{0pt}{3ex}       &            & 450.03/850.00  & 850.01 &   & 450.04/850.03 \\
\rule{0pt}{5ex}ALMA name$^{\rm{a}}$ & C2a                & C3a               & C14               & $\dots$              & C42 & \dots \\
\rule{0pt}{5ex}Redshift             & 3.179$^{\rm{i}}$   & 3.89$^{+3.11}_{-0.67}$\ $^{\rm{j}}$ & 4.58$^{+0.25}_{-0.68}$\ $^{\rm{l}}$ & 1.003$^{\rm{m}}$ & 3.63$^{+0.37}_{-0.56}$\ $^{\rm{l}}$ & \dots \\
\rule{0pt}{3ex}$S_{24\,\mu{\rm m}}^{\rm{\,a}}$ & $<\,$0.054 &  $<\,$0.181 & $<\,$0.054 & 0.26$\,\pm\,$0.02 & 0.189$\,\pm\,$0.013 & $<\,$0.054\\
\rule{0pt}{3ex}$S_{100\,\mu{\rm m}}^{\rm{\,a}}$& $<\,$7.7 & $<\,$5.0 & $<\,$5.0 & $<\,$6.9 & $<\,$5.0 & $<\,$8.5\\
\rule{0pt}{3ex}$S_{160\,\mu{\rm m}}^{\rm{\,a}}$& $<\,$33.5 & $<\,$15.6 & $<\,$10.2 & $<\,$36.1 & $<\,$10.2 & $<\,$17\\
\rule{0pt}{3ex}$S_{250\,\mu{\rm m}}^{\rm{\,a}}$& $<\,$62.7 & $<\,$33.3 & $<\,$14.4 & $<\,$47.6 & $<\,$54.1 & $<\,$14\\
\rule{0pt}{3ex}$S_{350\,\mu{\rm m}}^{\rm{\,a}}$& $<\,$62.0 & $<\,$47.1 & $<\,$22.9 & $<\,$41.4 & $<\,$57.5 & $<\,$19\\
\rule{0pt}{3ex}$S_{450\,\mu{\rm m}}^{\rm{\,b}}$& $\dots$ & $<\,$25.3 & $<\,$17.9 & 8.11$\,\pm\,$5.23 & 25.35$\,\pm\,$6.06 & $<\,$25\\
\rule{0pt}{3ex}$S_{500\,\mu{\rm m}}^{\rm{\,a}}$& $<\,$69.8 & $<\,$39.2 & $<\,$30.1 & $<\,$20.7 & $<\,$43.4 & $<\,$25\\
\rule{0pt}{3ex}$S_{850\,\mu{\rm m}}^{\rm{\,b,c}}$& 6.30$\,\pm\,$1.43$^{\rm{c}}$ & 11.73$\,\pm\,$1.08$^{\rm{b,k}}$ & 11.49$\,\pm\,$1.1$^{\rm{b}}$ &  2.12$\,\pm\,$0.98$^{\rm{b}}$ & 11.42$\,\pm\,$1.38$^{\rm{b}}$ & $<\,$5\\
\rule{0pt}{3ex}$S_{870\,\mu{\rm m}}^{\rm{\,d}}$& 12.3$\,\pm\,$3.6 & 11.47$\,\pm\,$2.39$^{\rm{k}}$ & 16.4$\,\pm\,$3.3 & $\dots$ & 7.2$\,\pm\,$0.2 & \dots\\
\rule{0pt}{3ex}$S_{890\,\mu{\rm m}}^{\rm{\,e}}$& 21.6$\,\pm\,$2.3 & 9.87$\,\pm\,$0.79 & 7.4$\,\pm\,$3.0 & $\dots$ & 9.3$\,\pm\,$1.3 & \dots\\
\rule{0pt}{3ex}$S_{1.1\,{\rm mm}}^{\rm{\,f,g}}$  & 5.5$\,\pm\,$1.3$^{\rm f}$ & 6.6$\,\pm\,$1.0$^{\rm f,k}$ & 5.8$\,\pm\,$1.3$^{\rm f}$ & $<\,$5.5$^{\rm g}$ & 6.5$\,\pm\,$1.2$^{\rm{f}}$ & $<\,$5.5$^{\rm g}$\\
\rule{0pt}{3ex}$S_{1.3\,{\rm mm}}^{\rm{\,a}}$  & 4.07$\,\pm\,$0.15 & 4.50$\,\pm\,$0.15 & 5.01$\,\pm\,$0.10 & $\dots$ & 2.39$\,\pm\,$0.1 & \dots\\
\rule{0pt}{3ex}$S_{2\,{\rm mm}}^{\rm{\,h}}$    & 1.09$\,\pm\,$0.25 & 0.68$\,\pm\,$0.21 & 0.75$\,\pm\,$0.26 & 0.50$\,\pm\,$0.32 & 0.50$\,\pm\,$0.32 & 0.70$\,\pm\,$0.27\\ [1ex]
\hline
\rule{0pt}{3ex} log($L_{\rm IR}/\,{\rm L}_\odot$) &  $12.6\pm0.3$ & $13.0\pm0.3$ & $13.0\pm0.2$ & $11.6\pm0.3$ & $13.1\pm0.1$ & \dots\\
\rule{0pt}{3ex} log($M_{\rm dust}/\,{\rm M}_\odot$) & $9.9\pm0.1$ & $9.6\pm0.1$ & $9.6\pm0.1$ & $9.2\pm0.3$ & $9.3\pm0.1$ & \dots\\
\rule{0pt}{3ex} log($M_{\ast} /\,{\rm M}_\odot$)$^{\rm{\,a, m}}$ & $10.97^{+0.01}_{-0.01}$\ $^{\rm{\,a}}$ & $\dots$ & $10.82^{+0.01}_{-0.10}$\ $^{\rm{\,a}}$  & $10.34^{+0.06}_{-0.08}$\ $^{\rm{\,m}}$ & $11.46^{+0.00}_{-0.00}$\ $^{\rm{\,a}}$ & \dots\\
\rule{0pt}{3ex} $t_{\rm depletion}$$^{\rm{\,n}}$ [Gyr] & $2.5^{+2.2}_{-1.3}$ & $0.6^{+0.6}_{-0.3}$ & $0.6^{+0.4}_{-0.2}$ & $5.7^{+7.3}_{-3.8}$ & $0.3^{+0.1}_{-0.1}$ & \dots\\ [1ex]
\hline
\end{tabular}
\end{center}
\textbf{Notes.}
All flux densities are in mJy.
\textit{Herschel} flux densities were all treated as upper limits, because of possible contamination by emission from low-redshift nearby galaxies.
$^{(\rm{a})}$ \citet{miettinen_2017c}.
$^{(\rm{b})}$ \citet{casey_2013}.
$^{(\rm{c})}$ \citet{geach_2017}.
$^{(\rm{d})}$ Navarrete et al. in prep.
$^{(\rm{e})}$ \citet{younger_2007,younger_2009}.
$^{(\rm{f})}$ \citet{scott_2008}.
$^{(\rm{g})}$ \citet{aretxaga_2011}.
$^{(\rm{h})}$ GISMO flux densities are defined as the average deboosted flux densities provided by the models of \citet{zavala_2018} and \citet{bethermin_2017}.
$^{(\rm{i})}$ Spectroscopic redshift reported in \citet{brisbin_2017}. No uncertainties are available for this redshift estimate.
$^{(\rm{j})}$ Photometric redshift reported in \citet{brisbin_2017} and inferred from 3-to-240\,GHz flux density ratio.
$^{(\rm{k})}$ Original flux densities have been scaled using the ALMA flux density ratio of the two components, i.e., 4.5/(4.5+1.15); see text for details.
$^{(\rm{l})}$ Photometric redshift reported in \citet{brisbin_2017} and inferred by fitting their optical-to-near-infrared photometry.
$^{(\rm{m})}$ Spectroscopic redshift reported in \citet{casey_2017}. No uncertainties are available for this redshift estimate.
$^{(\rm{n})}$ Gas depletion time defined as the ratio of the total gas mass to SFR, where the total gas mass is inferred assuming a gas-to-dust ratio of 100 and SFR\,[M$_{\odot}$\,yr$^{-1}$]$\,=\,10^{-10}\times L_{\rm IR}\,$[L$_{\odot}$] \citep{kennicutt_1998}.
$^{(\rm{o})}$ GISMO-C4, our fourth brightest source with a low false detection probability ($\sim\,$6.2\%), is within deep (sub)mm coverage of the COSMOS field but it has no counterpart in these surveys,  suggesting a high-redshift origin. To facilitate future follow-up studies, we summarize here current upper limits on its infrared-to-mm photometry.
\end{table*}

\acknowledgements
We would like to thank the referee for their comments which have helped improving the paper.
BM, AK, EJA and FB acknowledge support from the Collaborative Research Centre 956, sub-project A1, funded by the Deutsche Forschungsgemeinschaft (DFG). 
We would like to thank Carsten Kramer, Santiago Navarro, David John, Albrecht Sievers, and the entire IRAM Granada staff for their support during the instrument installation and observations.
CMC thanks the University of Texas at Austin College of Natural Sciences and NSF grants AST-1714528 and AST-1814034 for support.
D.R. acknowledges support from the National Science Foundation under grant number AST-1614213.
Based on observations carried out under project number 247-11, 227-12, 242-13, 117-14 and 232-15 with the IRAM 30-meter telescope. 
IRAM is supported by INSU/CNRS (France), MPG (Germany), and IGN (Spain). 
This work was also supported through NSF ATI grants 1020981 and 1106284.
\vspace{5mm}
\facilities{IRAM: 30m}

\end{document}